\documentclass[12pt]{article}

\usepackage{graphics}

\def\bfv{{\bf v}}
\def\bfu{{\bf u}}
\def\bfj{{\bf J}}

\begin{document}

\title{Characteristic Angles in the Wetting \\ of an Angular Region: Deposit Growth}
\author{Yuri O.\ Popov\thanks{Corresponding author.  E-mail: {\tt yopopov@midway.uchicago.edu}} \and Thomas A.\ Witten}
\date{{\em Department of Physics, University of Chicago,} \\
{\em 5640 S.\ Ellis Ave., Chicago, IL 60637}}
\maketitle


\begin{abstract}

As was shown in an earlier paper~\cite{deegan1}, solids dispersed in a drying drop migrate to the (pinned) contact line.  This migration is caused by outward flows driven by the loss of the solvent due to evaporation and by geometrical constraint that the drop maintains an equilibrium surface shape with a fixed boundary.  Here, in continuation of our earlier paper~\cite{popov}, we theoretically investigate the evaporation rate, the flow field and the rate of growth of the deposit patterns in a drop over an angular sector on a plane substrate.  Asymptotic power laws near the vertex (as distance to the vertex goes to zero) are obtained.  A hydrodynamic model of fluid flow near the singularity of the vertex is developed and the velocity field is obtained.  The rate of the deposit growth near the contact line is found in two time regimes.  The deposited mass falls off as a weak power $\gamma$ of distance close to the vertex and as a stronger power $\beta$ of distance further from the vertex.  The power $\gamma$ depends only slightly on the opening angle $\alpha$ and stays between roughly $-1/3$ and $0$.  The power $\beta$ varies from $-1$ to $0$ as the opening angle increases from 0 to 180 degrees.  At a given distance from the vertex, the deposited mass grows faster and faster with time, with the greatest increase in the growth rate occurring at the early stages of the drying process.

\end{abstract}

\begin{center}
\end{center}


\section{Introduction}

The problem of the so-called ``coffee-drop deposits'' has recently aroused great interest~\cite{deegan1, deegan2, deegan3, deegan4}.  The residue left when coffee dries on the countertop, mineral rings left on washed glassware, banded deposits of salt on the sidewalk during winter and enhanced edges in water color paintings are examples of the variety of physical systems understood by ``coffee-drop deposits'' terminology.  Understanding the process of drying of such solutions is important for paint manufacturing, protein crystallography, production of nanowires, patterning of a surface, DNA stretching in a flow, and other scientific and industrial applications.

The theory of the solute transfer in such physical systems has been developed in works of Deegan {\em et al.}~\cite{deegan1, deegan2, deegan3}.  In this theory, the contact line of a drop of liquid is pinned during drying process.  While the highest evaporation occurs at the edges, the bulk of the solvent is concentrated closer to the center of the drop.  In order to replenish the liquid removed by evaporation at the edge, a flow from the inner to the outer regions must exist inside the drop.  This flow is capable of transferring all of the solute to the contact line and thus accounts for the strong contact-line concentration of the residue left after complete drying.  The theory of Deegan {\em et al.}~\cite{deegan1} is very robust since it only requires the pinning of the edge during drying, which can occur in a number of possible ways (surface roughness, chemical heterogeneities {\em etc}), and it is independent of the nature of the solute.  It accounts quantitatively for the experimentally observed phenomena at least in the case of geometry analyzed in~\cite{deegan1}.  However, only the simplest case of a round drop was analytically solved by Deegan {\em et al.}

\begin{figure}
\begin{center}
\includegraphics{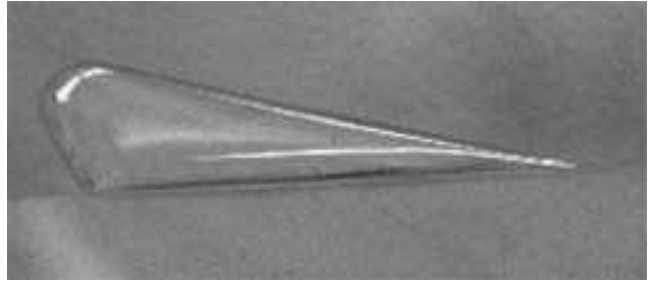}

(a)

\includegraphics{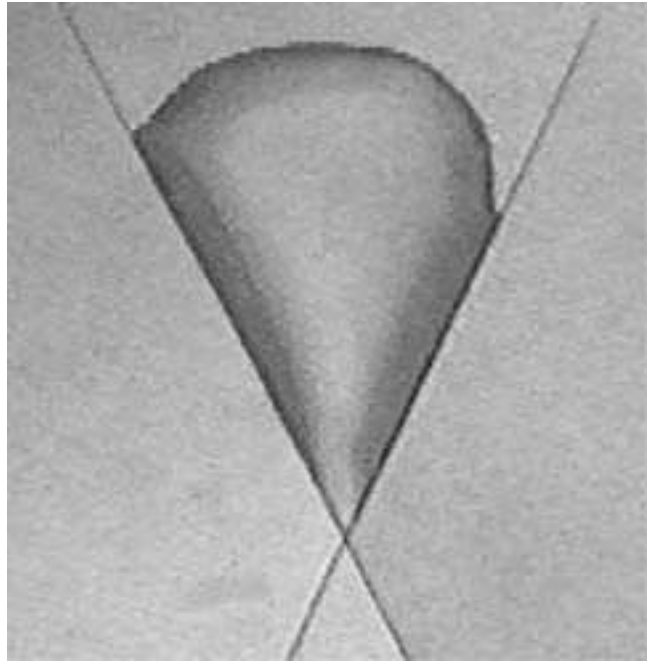}

(b)

\caption{(a) A water droplet with a sector-shaped boundary on the plane substrate (side view).  (b) The same droplet pictured from another point (top view).  Black lines are the grooves on the substrate necessary to ``pin'' the contact line.  (Photos by Itai Cohen.)}
\label{cohen}
\end{center}
\end{figure}

Here we consider a complementary problem of a solute drop drying over an angular region (Fig.~\ref{cohen}).  An arbitrary boundary line can be represented as a sequence of smooth segments, which can be approximated by circular arcs, and fractures, which can be approximated by angular regions.  Thus, knowledge of analytical solution for both circular (solved by Deegan {\em et al.}) and angular (considered here) boundary shapes fills out the quantitative picture of solute transfer and deposit growth for an arbitrary drop boundary.  Keeping this purpose in mind, we specify only one boundary of the drop (the vertex and the sides of the angle) leaving the remainder of the boundary curve unspecified.  Such approach turns out to be sufficient to determine the universal features of the solution, and it allows us to find all the important singularities as power laws of distance from the vertex of the angle.

\begin{figure}
\begin{center}
\includegraphics{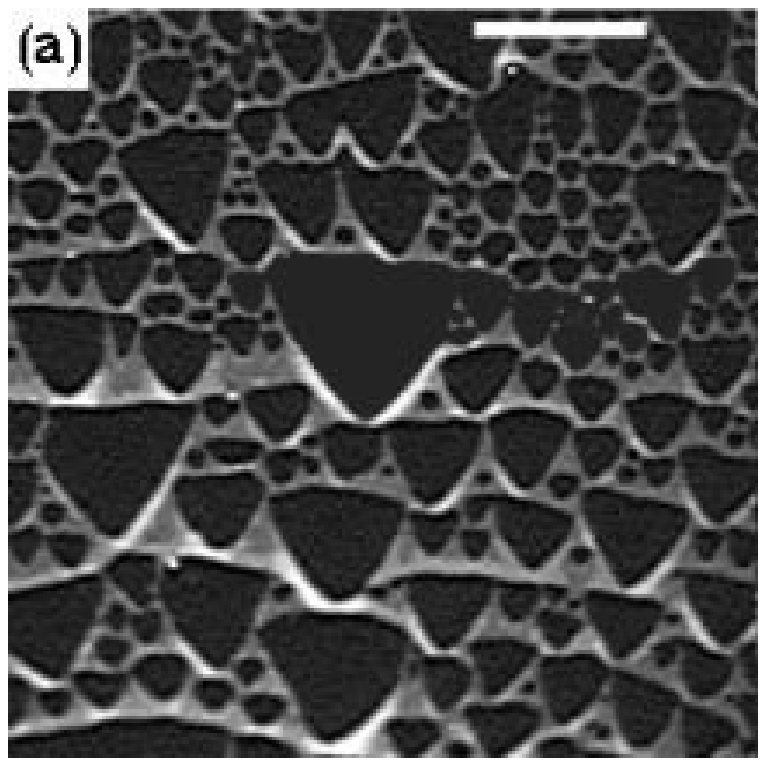}
\caption{Deposit patterns observed in experiments by Deegan {\em at al.}  These patterns were left after the contact line moved through the area shown.  The contact line was retreating down on this image and the solute was below the contact line.  (Photo by Robert Deegan.)}
\label{deegan}
\end{center}
\end{figure}

Our motivation for focusing on droplets over an angular sector also arises from the distinctive deposit patterns observed in Deegan's experiments (Fig.~\ref{deegan}).  These patterns are produced as the contact line retracts (down on the figure) discontinuously, switching between pinned and moving states.  We expect that the knowledge of the solution for a single pinned angle can boost understanding of these distinctive dynamically-produced patterns.

The flow pattern explored in Deegan's work~\cite{deegan1, deegan2} is a new type of capillary flow that depends sensitively on the perimeter shape.  The range of flows and deposition profiles obtainable by this mechanism remains virtually unexplored.  Our study aims to explore the extremes of what behavior can be obtained by varying the perimeter shape.

As a result of our study, we find that the solution in the angular-sector geometry is in a number of ways different from the solution in the circular geometry.  In particular, three time regimes for the deposit growth are found in the angular case compared to the two regimes in the circular one.  A clear-cut signature of the existence of these new regimes is also provided.  In addition, the new geometry possesses an extra free parameter --- the opening angle of the sector, --- and dependence of all the universal exponents in the power laws for all quantities on this extra parameter yields the possibility to control the deposition patterns by simply choosing a proper opening angle.  Thus, we obtain results that are both universal ({\em i.e.}\ do not depend on any physical properties of the constituents) and at the same time dependent on the geometry in a predictable fashion.  These and other useful properties of our results are discussed at the end of this manuscript.

In the following section we first consider the physical assumptions and the mathematical formulation of our theory, then calculate the equilibrium surface shape of the drop and the rate of evaporation from the surface of the drop, and finally obtain the velocity field inside the drop.  At the end of the section we describe the trajectories of the particles and provide the results for the solute transfer to the contact line and the deposit growth in different time regimes.  In the last section we discuss achievements and limitations of our theory and explore the avenues for further study.

\section{Theory}

\subparagraph{System and assumptions.}  We consider a droplet of solution on a horizontal surface bounded by an angle $\alpha$ in the plane of the substrate (Fig.~\ref{cohen}).  We assume that the droplet is sufficiently small so that the surface tension is dominant, and the gravitational effects can be safely neglected.  At the same time, we do {\em not\/} assume that the contact angle between the liquid-gas interface and the plane is constant along the boundary line on the substrate, nor do we assume it is constant in time.  To achieve an angular boundary, the substrate must have scratches, grooves or other inhomogeneities (sufficiently small compared to the dimensions of the droplet), which {\em pin\/} the contact line.  A strongly pinned contact line can sustain a wide range of contact angles; the angle is not fixed by the interfacial tensions as it is on a uniform surface.

\begin{figure}
\begin{center}
\includegraphics{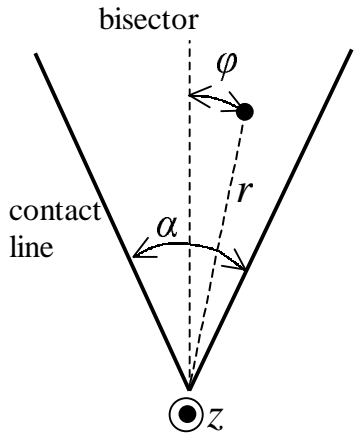}
\caption{Geometry of the problem.  The plane of the figure coincides with the substrate.}
\label{geometry}
\end{center}
\end{figure}

The use of the cylindrical coordinates $(r,\phi,z)$ is most natural in this problem, so that the angle occupied by the liquid on the substrate is $0<r<\infty$ and $-\alpha/2<\phi<\alpha/2$, and the coordinate normal to the substrate is $z$ (Fig.~\ref{geometry}).  The geometry of this problem is much more complicated than the one of the round-drop case solved earlier.  We consider a two-dimensional object --- the angular sector --- in a three-dimensional space, and the main complication arises from the fact that the symmetry of the object does not match the symmetry of any simple orthogonal coordinate system in that space.  In particular, even the solution of the Laplace equation (needed below) requires introduction of the special coordinate system (the so-called conical coordinates, or the orthogonal coordinates of the elliptic cone) with heavy use of the Jacobi elliptic functions.  Similarly, a separate research~\cite{popov} was required to find the equilibrium surface shape of the drop in this geometry.  Thus, given this complex geometry, we limit our task to determining only the power-law scaling for most quantities, and this task proves to be sufficiently challenging by itself.

We describe the surface shape of the drop $h(r,\phi)$ by local mean curvature $H$ that is spatially uniform at any given moment of time, but changes with time as droplet dries.  Ideally, the surface shape should be considered dynamically together with the flow field inside the drop.  However, as we show in the Appendix, for flow velocities much lower than the characteristic velocity $v^* = \sigma/3\eta$ ($\sigma$ is surface tension and $\eta$ is dynamic viscosity), which is about 24~m/s for water under normal conditions, one can consider the surface shape independently of the flow and use the equilibrium result at any given moment of time for finding the flow at that time.  The equilibrium surface shape $h(r,\phi)$ for the drop over an angular region was found in our earlier paper~\cite{popov}.

\begin{figure}
\begin{center}
\includegraphics{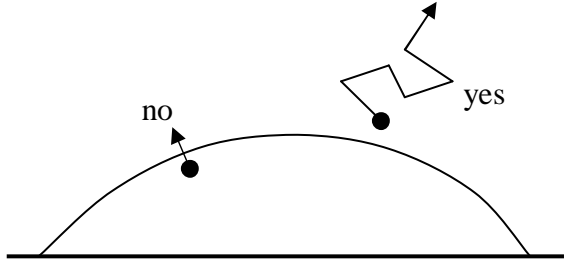}
\caption{The rate-limiting process for evaporative mass loss.  It is the diffusion of saturated vapor just above the interface rather than the transfer across the interface.}
\label{rlp}
\end{center}
\end{figure}

In order to determine the flow caused by evaporation, one needs to know the flux profile of liquid leaving each point of the surface by evaporation.  The functional form of the evaporation rate $J(r,\phi)$ (defined as evaporative mass loss per unit surface area per unit time) depends on the rate-limiting step, which can, in principle, be either the transfer rate across the liquid-vapor interface or the diffusive relaxation of the saturated vapor layer immediately above the drop.  As in the work of Deegan {\em et al.}, we assume that the rate-limiting step is diffusion of liquid vapor (Fig.~\ref{rlp}) and that evaporation rapidly attains a steady state.  Indeed, the transfer rate across the liquid-vapor interface is characterized by the time scale of the order of $10^{-10}$~s, while the diffusion process has characteristic times of the order of $R^2/D$ (where $D$ is the diffusion constant for vapor in air and $R$ is a characteristic size of the drop), which is of the order of seconds for water drops under typical drying conditions.

We consider {\em slow\/} flows, {\em i.e.}\ flows with low Reynolds numbers (also known as ``creeping flows'').  This amounts to the neglect of the inertial terms in the Navier-Stokes equation.  We also employ the so-called ``lubrication approximation''.  It is essentially based on the two conditions reflecting the thinness of the drop and resulting from the separation of the vertical and horizontal scales.  One is that the pressure inside the drop $p$ does not depend on the coordinate $z$ normal to the substrate:  $\partial_z p = 0$.  The other is related to the small slope of the free surface $|\nabla h| \ll 1$, which is equivalent to the dominance of the $z$-derivatives of any component $u_i$ of flow velocity $\bfu$: $\partial_z u_i \gg \partial_s u_i$ (index $s$ refers to the derivatives with respect to any coordinate in the plane of the substrate).  The lubrication approximation is a standard simplifying procedure for this class of hydrodynamic problems~\cite{greenspan, cameron, brenner}.

Having formulated physical assumptions intrinsic to the theory, we are now in position to formulate its main ideas.

\subparagraph{Equations.}  The essential idea behind the theory is that a pinned contact line entails fluid flow toward that contact line, since the rate of evaporation is the highest at the edge of the drop while the most of the liquid is concentrated away from it.  The ``elasticity'' of the liquid-air interface fixed at the contact line provides the force driving this flow.

\begin{figure}
\begin{center}
\includegraphics{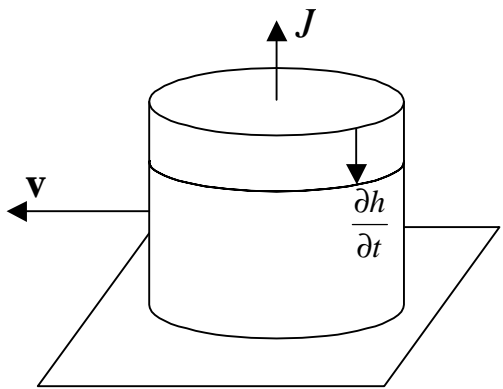}
\caption{Conservation of mass: the liquid-vapor interface lowers exactly by the amount of fluid evaporated from the surface plus the difference between the outflow and the influx of fluid from the adjacent regions.}
\label{consmasseps}
\end{center}
\end{figure}

To develop this idea mathematically, we ignore for a moment any solute in the liquid.  Once the flow is found, one can track the motion of the suspended particles, since they are just carried along by the flow.  We define depth-averaged flow velocity by
\begin{equation}
\bfv = \frac 1h \int_0^h \bfu_s \, dz,
\label{defv}\end{equation} 
where $\bfu_s$ is the in-plane component of the local three-dimensional velocity $\bfu$.  Then we write the conservation of fluid mass in the form
\begin{equation}
\nabla\cdot(h\bfv) + \frac{J}{\rho}\sqrt{1+(\nabla h)^2} + \partial_t h = 0,
\label{consmass}\end{equation}
where $t$ is the time, $\rho$ is the density of the fluid and quantities $h$, $J$ and $\bfv$ are functions of $r$, $\phi$ and $t$.  (We will drop the $\nabla h$ part of the second term everywhere in the following since it is always small compared to unity, as it will be seen from the expression for $h(r,\phi)$ below.)  This equation represents the fact that the rate of change of the amount of fluid in a volume element (column) above an infinitesimal area on the substrate (third term) is equal to the negative of the sum of the net flux of liquid out of the column (first term) and the amount of mass evaporated from the surface element on top of that column (second term); Fig.~\ref{consmasseps} illustrates this idea.  Thus, this expression relates the depth-averaged velocity field $\bfv (r,\phi,t)$ to the liquid-vapor interface position $h(r,\phi,t)$ and the evaporation rate $J(r,\phi,t)$.  However, this is only one equation for two variables since vector $\bfv$ has two components in the plane of the substrate.  Moreover, as it was pointed out earlier, while the evaporation rate $J$ is indeed independent of flow $\bfv$, the free-surface shape $h$ should in general be determined simultaneously with $\bfv$.  Thus, there are actually three unknowns to be determined together ($h$, $v_r$ and $v_\phi$), and hence two more equations are needed.  In order to find these equations we will employ the lubrication approximation.

We start with the Navier-Stokes equation with inertial terms omitted (low Reynolds numbers):
\begin{equation}
\nabla p = \eta \nabla^2 \bfu,
\end{equation}
where $p$ is the fluid pressure, $\eta$ is the dynamic viscosity and $\bfu$ is the velocity.  Applying lubrication-approximation conditions $\partial_z p = 0$ and $\partial_z u_i \gg \partial_s u_i$, we arrive at the simplified form of this equation
\begin{equation}
\nabla_s\,p = \eta\,\partial_{zz} \bfu_s,
\end{equation}
where index $s$ again refers to the vector components along the substrate.  From now on we will suppress the subscript $s$ at the symbol of nabla-operator, and will assume for the rest of this paper that this operator refers to the two-dimensional vector operations in the plane of the substrate.  Solution to the above equation with boundary conditions
\begin{equation}
\left. \bfu_s \right|_{z = 0} = 0 \qquad\qquad\mbox{and}\qquad\qquad \left. \partial_z \bfu_s \right|_{z = h} = 0
\end{equation}
yields
\begin{equation}
\bfu_s = \frac{\nabla p}{\eta} \left(\frac{z^2}2 - h z\right),
\label{velocity-profile}\end{equation}
or, after vertical averaging~(\ref{defv}),
\begin{equation}
\bfv = - \frac{h^2}{3\eta} \nabla p.
\label{darcy}\end{equation}
This result is a variant of the Darcy's law~\cite{brenner, bensimon}.  Note that since a curl of a gradient is always zero and $\eta$ is a constant, the previous equation can be re-written as
\begin{equation}
\nabla\times\left(\frac{\bfv}{h^2}\right) = 0.
\label{curl}\end{equation}
This condition is analogous to the condition of the potential flow ($\nabla\times\bfv = 0$), but with a quite unusual combination of the velocity and the surface height $\bfv/h^2$ in place of the usual velocity $\bfv$.

Relation~(\ref{darcy}) provides the two sought equations in addition to the conservation of mass~(\ref{consmass}).  However, it contains one new variable, pressure $p$, and hence another equation is needed.  This last equation is provided by the condition of the mechanical equilibrium of the liquid-air interface (also known as the Young-Laplace equation) relating the pressure and the surface shape:
\begin{equation}
p = - 2 H \sigma + p_{atm}.
\label{mechequil}\end{equation}
Here $p_{atm}$ is the atmospheric pressure, $\sigma$ is the surface tension and $H$ is the mean curvature of the surface, uniquely related to the surface shape $h$ by differential geometry.  Note that this expression is independent of both the conservation of mass~(\ref{consmass}) and the Darcy's law~(\ref{darcy}).  Thus, the complete set of equations required to fully determine the four dynamic variables $h$, $p$, $v_r$ and $v_\phi$ consists of four differential equations (together with the appropriate boundary conditions at the contact line):  one equation of the conservation of mass~(\ref{consmass}), two equations of the Darcy's law~(\ref{darcy}), and one equation of the mechanical equilibrium of the interface~(\ref{mechequil}).  They provide all the necessary conditions to solve the problem at least in principle.

In practice, however, solution of these four {\em coupled\/} differential equations is not possible in the geometry of interest.  At the same time, under normal drying conditions the viscous stress is negligible, or, equivalently, the typical velocities are much smaller than $v^* = \sigma/3\eta \approx 24\mbox{ m}/\mbox{s}$ (for water).  As shown in the Appendix, the four equations {\em decouple\/} under these conditions.  As a result, one can employ the equilibrium result for the surface shape $h(r,\phi)$ at any given moment of time, and then determine the pressure and the velocity fields for this fixed functional form of $h$.  Mathematically, the original system of equations can be rewritten as:
\begin{equation}
2 H = - \frac{\Delta p}\sigma,
\label{laplace}\end{equation}
\begin{equation}
\nabla\cdot(h^3 \nabla\psi) = - \frac{J}{\rho} - \partial_t h,
\label{psi}\end{equation}
\begin{equation}
\bfv = h^2 \nabla\psi,
\label{vpsi}\end{equation}
where $\Delta p = p_0 - p_{atm}$, $\psi = - \epsilon p_1 / 3\eta$, and $p_0$ and $p_1$ are the leading and the first-order terms in the expansion of pressure $p = p_0 + \epsilon p_1 + \cdots$ in a small parameter $\epsilon$ inversely proportional to $v^*$ (see the Appendix for details).  Note that $p_0$ is independent of $(r,\phi)$, although it does depend on time (and this time dependence will be determined later in this paper).  Therefore, there is a profound difference between equations~(\ref{mechequil}) and (\ref{laplace}):  the former is a local statement, with the right-hand side depending on the coordinates of a point within the angular sector, while the latter is a global condition of spatial constancy of the mean curvature throughout the drop.  Equation~(\ref{laplace}) defines the {\em equilibrium\/} surface shape for any given value of $p_0$ at any given moment of time, and moreover, can be solved independently of the other equations.  Thus, the procedure for finding the solution becomes significantly simplified:  first find the equilibrium surface shape $h(r,\phi)$ from condition~(\ref{laplace}) and independently specify the functional form of the evaporation rate $J(r,\phi)$, then solve equation~(\ref{psi}) for the reduced pressure $\psi(r,\phi)$, and finally obtain the flow field $\bfv(r,\phi)$ according to prescription~(\ref{vpsi}).  The next three sections will be devoted to the particular steps of this procedure.

\subparagraph{Surface shape.}  There are well defined equations governing the equilibrium surface shape $h(r,\phi)$ and the evaporation rate $J(r,\phi)$, but there is no generic method for solving these equations analytically in an arbitrary geometry, in particular, in the geometry of the angular sector under consideration.  Moreover, even if one could find these exact analytic expressions, a second-order differential equation of the kind of Eq.~(\ref{psi}) would not be guaranteed to have an analytic solution in closed form for arbitrary functions $h$ and $J$.  Hence, a feasible way to proceed analytically is to seek an approximate solution that captures the essential physical features and is correct at least asymptotically.  Here, in the geometry of an angular sector, the only possible locations of singularities and divergences (which normally govern the properties of the solution) are at the vertex of the angle ({\em i.e.}\ at $r = 0$) and at its sides ({\em i.e.}\ at $\phi = \pm \alpha/2$).  Therefore, the most important physical features will be correctly reflected if asymptotic results (as $r \to 0$ and as $\phi \to \pm \alpha/2$) are found analytically.

\begin{figure}
\begin{center}
\includegraphics{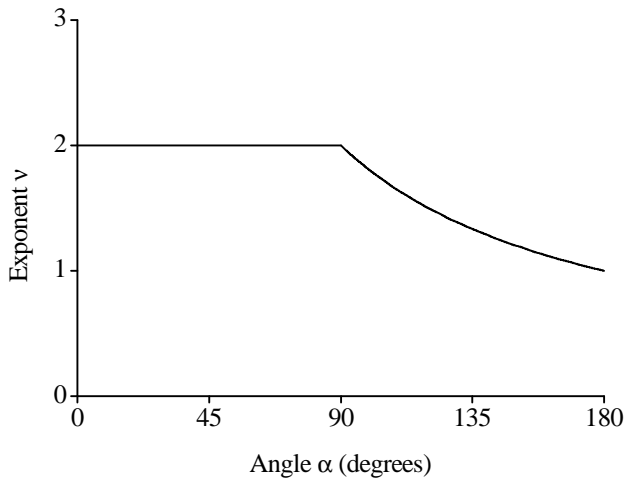}
\caption{Dependence of exponent $\nu$ in the power law $h(r)$ [Eq.~(\ref{surfshap})] on opening angle $\alpha$, after Ref.~\cite{popov}.}
\label{nueps}
\end{center}
\end{figure}

The boundary problem for the equilibrium surface shape of the drop consists of the differential equation~(\ref{laplace}) and boundary conditions at the vertex and at the sides of the angle:
\begin{equation} h(0, \phi) = h(r, -\alpha/2) = h(r,\alpha/2) = 0.
\label{boundary} \end{equation}
Equation~(\ref{laplace}) represents the fact that the local mean curvature is spatially uniform, but changes with time as the right-hand side ($\Delta p$) changes during the drying process.  The asymptotic solution to the boundary problem~(\ref{laplace}), (\ref{boundary}) was found in our earlier paper~\cite{popov}.  The result turned out to have two qualitatively different regimes in opening angle $\alpha$ (acute and obtuse angles) and can be written as
\begin{equation}
h(r,\phi) = \frac{r^\nu \tilde h(\phi)}{R^{\nu-1}}.
\label{surfshap}\end{equation}
Here $R(t) = \sigma/\Delta p$ and is the only function of time in this expression; exponent $\nu$ has a discontinuous derivative at $\alpha = \pi/2$ and is shown in Fig.~\ref{nueps}; and
\begin{equation}
\tilde h(\phi) = \left\{\begin{array}{lll} \frac 14 \left(\frac{\cos 2\phi}{\cos\alpha}-1\right) &\qquad\qquad\mbox{if}\quad 0\le\alpha<\frac \pi 2 &\qquad(\nu=2),\\ \\ C \cos\frac{\pi\phi}{\alpha} &\qquad\qquad\mbox{if}\quad \frac \pi 2 <\alpha\le\pi &\qquad(\nu=\pi/\alpha). \end{array}\right.
\label{phi}\end{equation}
The constant $C$ cannot be determined without imposing boundary conditions on $h$ at some curve on the side of the drop furthest from the vertex of the angle.  It is restricted by neither the equation nor the side boundary conditions, and thus, is not a universal feature of the solution near the vertex of the angle.  The constant $C$ can (and does) depend on the opening angle $\alpha$.  As we showed in the earlier paper, this constant must have the following diverging form near $\alpha = \pi/2$:
\begin{equation}
C = \frac 1{4\alpha - 2\pi} + C_0 + O(\alpha - \pi/2)
\label{c-choice}\end{equation}
where $C_0$ is independent of $\alpha$.  We will adopt this form of $C$ (with $C_0$ set to unity) for all numerical estimates for obtuse opening angles.

Two different values of $\nu$ corresponding to the acute and obtuse angles give rise to the two qualitatively different regimes for surface shape.  This difference can best be seen from the fact that the principal curvatures of the surface stay finite as $r \to 0$ for acute angles and diverge as a power of $r$ for obtuse angles.  This qualitative difference can be observed in a simple experimental demonstration, which we provided in our earlier work~\cite{popov}.  We refer to that earlier work for further details and discussion.  We only note here that the asymptotic $r \to 0$ at the vertex of the angle actually means $r \ll R$ (which is typically of the order of a few millimeters for water under normal conditions), and that $\nabla h$ is indeed small for $r \ll R$ and can be safely neglected with respect to unity ({\em i.e.}\ the free surface is nearly horizontal in the vicinity of the tip of the angle) as was asserted earlier.

\subparagraph{Evaporation rate.}  Since the rate-limiting step in the evaporation process is diffusion of saturated vapor just above the liquid-vapor interface, density of vapor $n$ obeys the diffusion equation.  However, diffusion rapidly attains a steady state (typically in a fraction of a second), and therefore the diffusion equation reduces to the Laplace equation
\begin{equation}
\nabla^2 n = 0.
\end{equation}
This equation is to be solved together with the following boundary conditions: (a) along the surface of the drop the air is saturated with vapor and hence $n$ at the interface is the constant density of saturated vapor $n_s$, (b) far away from the drop the density approaches the constant ambient vapor density $n_\infty$, and (c) vapor cannot penetrate the substrate and hence $\partial_z n = 0$ at the substrate outside of the drop.  Having found density of vapor, one can obtain the evaporation rate $\bfj = - D \nabla n$, where $D$ is the diffusion coefficient.

\begin{figure}
\begin{center}
\includegraphics{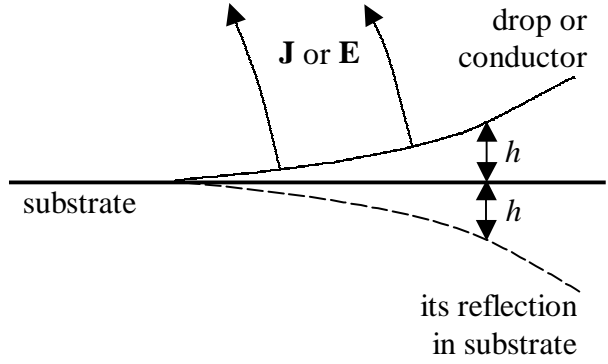}

\caption{Illustration of analogy between evaporation rate {\bf J} for a liquid drop and electric field {\bf E} for a conductor.  Consideration of the drop (or conductor) and its reflection in the plane of the substrate significantly simplifies the boundary problem.}
\label{evaprateeps}
\end{center}
\end{figure}

This boundary problem is mathematically equivalent to that of a charged conductor of the same geometry at constant potential if we identify $n$ with the electrostatic potential and $\bfj$ with the electric field.  Moreover, since there is no component of $\bfj$ normal to the substrate, we can further simplify the boundary problem by considering a conductor of the shape of our drop plus its reflection in the plane of the substrate in the full space instead of viewing only the semi-infinite space bounded by the substrate (Fig.~\ref{evaprateeps}).  This reduces the number of boundary conditions to only two: (a) $n = n_s$ on the surface of the conductor, and (b) $n = n_\infty$ at infinity.  The shape of the conductor (the drop over the angular sector and its reflection in the substrate) now resembles a dagger blade.  So, now we have to tackle the problem of finding the electric field around the tip of a dagger blade at constant potential in infinite space.

If one decides to account for the thickness of the blade (given by doubled $h(r,\phi)$ of Eq.~(\ref{surfshap})) accurately, it becomes apparent that there is no hope for any analytical solution in this complex geometry.  However, taking into account that near the tip $\nabla h$ is very small and hence the thickness of the blade itself is very small, we can approximate our thick dagger blade with a dagger blade of zero thickness and the same opening angle ({\em i.e.}\ with a flat angular sector).  In the limit $r \to 0$ the contact angle $\theta$ scales with $r$ as $(r/R)^{\nu-1}$ and hence goes to zero.  Thus, only the flat blade can be considered up to the main order in $r$.  This approximation would not be adequate for determining the surface shape or the flow field, but it is perfectly adequate for finding the evaporation rate.  We will discuss possible corrections to this result later in this subsection.

\begin{figure}
\begin{center}
\includegraphics{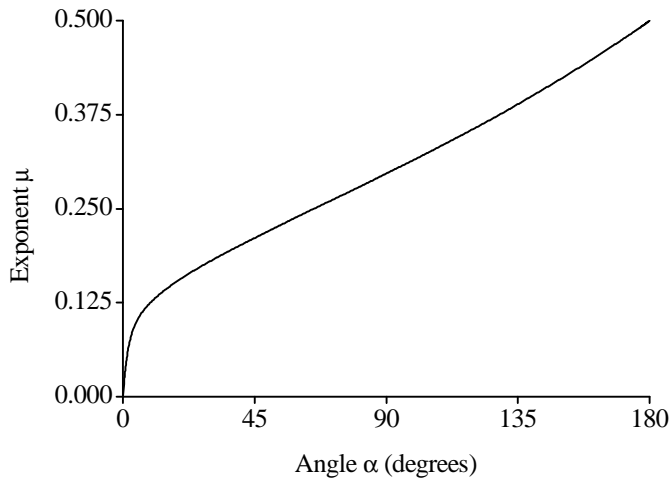}
\caption{Dependence of exponent $\mu$ in the power law $J(r)$ [Eqs.~(\ref{j-def}) and (\ref{evaprate})] on opening angle $\alpha$, after Refs.~\cite{blume1, blume2, desmedt}.}
\label{mueps}
\end{center}
\end{figure}

The problem of finding the electric field and the potential for an infinitely thin angular sector in three-dimensional space requires introduction of the so-called conical coordinates (the orthogonal coordinates of the elliptic cone) and heavily involves various special functions.  Luckily, it was studied extensively in the past~\cite{kraus, blume1, blume2, desmedt}, although the results cannot be expressed in a closed form.  An important conclusion from these studies is that the $r$ and $\phi$ dependences separate and that the electric field near the vertex of the sector scales with $r$ as a power law with an exponent depending on the opening angle $\alpha$:
\begin{equation}
J \propto r^{\mu - 1} \tilde J(\phi).
\label{j-def}\end{equation}
Here
\begin{equation}
\tilde J(\phi) \propto \frac 1{|\cos\phi^*|} \left.\frac{\partial Y_\mu(\theta^*,\phi^*)}{\partial\theta^*}\right|_{\theta^* = \pi},
\label{theta-def}\end{equation}
and $\mu$ and $Y_\mu(\theta^*,\phi^*)$ are the eigenvalue and the eigenfunction, respectively, of the eigenvalue problem
\begin{equation}
- {\bf L}^2 Y_\mu(\theta^*,\phi^*) = \mu (\mu + 1) Y_\mu(\theta^*,\phi^*)
\label{eigenvalue}\end{equation}
with Dirichlet boundary conditions on the surface of an elliptic cone (degenerating to an angular sector as $\theta^* \to \pi$).  In the last relation ${\bf L}^2$ is the angular part of the Laplacian in conical coordinates ($r$, $\theta^*$, $\phi^*$).  On the surface of the sector ({\em i.e.}\ at $\theta^* = \pi$) the relation between the conical coordinate $\phi^*$ and the usual polar coordinate $\phi$ is $\sin\phi = \sin(\alpha/2) \sin\phi^*$.  We refer to work~\cite{blume2} for further details.  Here we notice only that neither $\mu$ nor $Y_\mu(\theta^*,\phi^*)$ can be expressed in a closed analytic form; however, the exponent $\mu$ can be computed numerically and is shown in Fig.~\ref{mueps} as a function of $\alpha$.  Note that this exponent is {\em lower\/} than similar exponents for corresponding angles for both a wedge (a two-dimensional corner with an infinite third dimension) and a circular cone.  Both these shapes (wedge and cone) allow simple analytical solutions but none of them would be appropriate for the zero-thickness sector.

Despite the unavailability of an explicit analytical expression for $\tilde J(\phi)$, its analytic properties at $\phi = 0$ and at $\phi = \pm \alpha/2$ are quite straightforward to infer.  Indeed, $\tilde J(\phi)$ is an even function of $\phi$; therefore, $\tilde J'(0) = 0$ (as well as any other odd derivative on the bisector) and
\begin{equation}
\tilde J(\phi) = \tilde J(0) + O(\phi^2)
\label{theta-bisector}\end{equation}
for small $\phi$.  Obviously, $\tilde J(0)$ is positive.  On the other hand, at $\phi = \pm \alpha/2$ the leading asymptotic of the evaporation rate (or the electric field) is known to be $(\Delta\phi)^{-1/2}$ with exponent $-1/2$ corresponding to the edge of an infinitely thin half plane in the three-dimensional space~\cite{jackson}.  (We have introduced notation $\Delta\phi = \alpha/2 - |\phi|$ in the previous line.)  If one were to correct this asymptotic in order to reflect the non-zero contact angle at the edge of the sector, the asymptotic form at $\phi = \pm \alpha/2$ would have to be written as
\begin{equation}
\tilde J(\phi) \approx J^* (\Delta\phi)^{-\lambda}
\label{theta-edge}\end{equation}
where $J^*$ is a positive constant and
\begin{equation}
\lambda = \frac{\pi-2\theta}{2\pi-2\theta}.
\label{lambda}\end{equation}
This result corresponds to the divergence of the electric field along the edge of a wedge of opening angle $2\theta$ (both the drop and its reflection contribute to the opening angle, hence a factor of 2)~\cite{jackson}.  However, accounting for the non-zero $\theta$ is a first-order correction to the main-order result $\lambda = 1/2$.  This can be seen from the expression for the contact angle:
\begin{equation}
\theta = \arctan\left[\left(\frac{r}{R}\right)^{\nu-1} \left|\tilde h'\left(\alpha/2\right)\right|\right] \propto \left(\frac{r}R\right)^{\nu-1}.
\label{contactangle}\end{equation}
For all opening angles $\nu > 1$ (except $\alpha = \pi$ where $\nu = 1$).  Thus, the correction due to the non-zero contact angle can indeed be neglected in the main-order results, and $\lambda$ should indeed be set to $1/2$.  Nevertheless, we will keep the generic notation $\lambda$ for this exponent in order to keep track of the origin of different parts of the final result and in order to account properly for the case $\alpha = \pi$ in addition to the range of opening angles below $\pi$.  The numerical value of $\lambda$ will be assumed to be $1/2$ in all estimates.

For the purposes of the numerical estimates {\em only\/} we will employ the following simplified form of $\tilde J(\phi)$:
\begin{equation}
\tilde J(\phi) = \left(\left(\frac\alpha2\right)^2 - \phi^2\right)^{-1/2}.
\label{model-2}\end{equation}
This model form satisfies both asymptotics~(\ref{theta-bisector}) and (\ref{theta-edge}), and it allows one to avoid the numerical solution of the eigenvalue problem~(\ref{eigenvalue}) and thus not to repeat the elaborate treatment of works~\cite{kraus, blume1, blume2, desmedt}.  This form is neither exact nor the only one satisfying the asymptotics, and the numerical graphs based on this form should be taken with a pinch of salt.  However we expect it to be a good approximation to the true function $\tilde J(\phi)$, and in order to check this we conducted all the numerical calculations for an alternative form of $\tilde J(\phi)$ as well:
\begin{equation}
\tilde J(\phi) = \left(\left(\frac\alpha2\right)^4 - \phi^4\right)^{-1/2}.
\label{model-4}\end{equation}
This alternative form also satisfies both asymptotics~(\ref{theta-bisector}) and (\ref{theta-edge}).  In most cases (all but one) the discrepancy between the numerical results based on the two model forms did not exceed 20\%, and this fact has convinced us that at least the orders of magnitude obtained by this approximation are correct.  We would like to emphasize that only the {\em numerical\/} graphs based on the choice of $\tilde J(\phi)$ are affected by these simplified forms; all the {\em analytical\/} results below do not rely on a particular form of $\tilde J(\phi)$ and use only the analytical asymptotics of the previous paragraph.

Thus, we will use the following expression for the evaporation rate $J$:
\begin{equation}
J(r,\phi) = J_0 \left(\frac{r}{\sqrt{A}}\right)^{\mu-1} \tilde J(\phi),
\label{evaprate}\end{equation}
where function $\tilde J(\phi)$ is defined in Eq.~(\ref{theta-def}) with asymptotics~(\ref{theta-bisector}) and (\ref{theta-edge}).  Here we broke down the constant pre-factor into two pieces:  a distance scale $\sqrt{A}$ (where $A$ is the substrate area occupied by the drop) and all the rest $J_0$ (which is of dimensionality of the evaporation rate).  Trivially, $J_0$ is directly proportional to the difference of the saturated and the ambient vapor densities $(n_s - n_\infty)$.

The evaporation rate does not depend on time and the same form of $J$ applies during the entire drying process, as the diffusion process is steady.  The same is true for the total rate of mass loss $dM/dt$ since
\begin{equation}
\frac{dM}{dt} = - \int_A J \sqrt{1+(\nabla h)^2} \, r dr d\phi \approx - \int_A J \, r dr d\phi \propto - J_0 A,
\label{dmdt}\end{equation}
where the integrations are over the substrate area occupied by the drop.  The constancy of this rate during most of the drying process was also confirmed experimentally~\cite{deegan3}.  This fact can be used to determine the time dependence of the length scale $R$ of Eq.~(\ref{surfshap}) (and hence of the pressure $p_0$ of Eq.~(\ref{laplace})) explicitly, as the mass $M$ of a sufficiently thin drop is inversely proportional to the mean radius of curvature $R$:
\begin{equation}
M \propto \frac{\rho A^2}R,
\label{totalmass}\end{equation}
where we retained only dimensional quantities and suppressed all the numerical pre-factors sensitive to the details of the drop shape.  From the last two equations one can conclude that
\begin{equation}
\frac{d}{dt}\left(\frac 1R\right) \propto - \frac{J_0}{\rho A}
\end{equation}
and remains constant during most of the drying process.  Hence,
\begin{equation}
R(t) = \frac{R_i}{1-t/t_f},
\label{r-t}\end{equation}
where $R_i$ is the initial mean radius of curvature ($R_i = R(0)$) and $t_f$ is the total drying time:
\begin{equation}
t_f \propto \frac{\rho A}{J_0 R_i}.
\label{tf}\end{equation}
Thus, at early drying stages ($t \ll t_f$) scale $R$ grows linearly with time; this time dependence will be implicitly present in the results below.  However, it is very weak at sufficiently early times and will be occasionally ignored (by setting $R \approx R_i$) when only the main-order results are of interest.

\begin{figure}
\begin{center}
\includegraphics{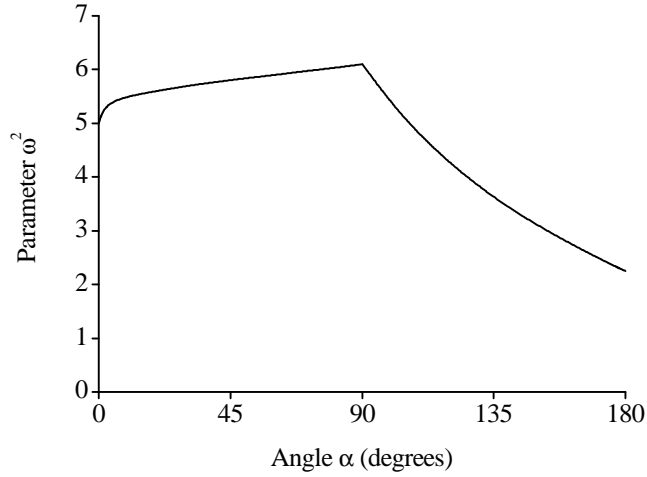}
\caption{Dependence of parameter $\omega$ of Eq.~(\ref{omega}) [governing solution $\tilde\psi(\phi)$ of Eq.~(\ref{psi-equation})] on opening angle $\alpha$.}
\label{omegaeps}
\end{center}
\end{figure}

\subparagraph{Flow field.}  With $h$ and $J$ in hand we proceed by solving Eq.~(\ref{psi}) for the reduced pressure $\psi$.  Assuming power-law divergence of $\psi$ as $r \to 0$ and leaving only the main asymptotic (which effectively means that we neglect the regular term $\partial_t h$ with respect to the divergent one $J/\rho$), we arrive at the following asymptotically-correct expression for $\psi$:
\begin{equation}
\psi(r,\phi) = \frac{J_0}\rho \frac{r^{\mu - 3\nu + 1}}{\sqrt{A}^{\mu - 1} R^{- 3\nu + 3}} \tilde\psi(\phi),
\label{psi-result}\end{equation}
where time-dependence is implicitly present via $R$ and the function $\tilde\psi(\phi)$ is a solution to the following differential equation:
\begin{equation}
\tilde\psi^{\prime\prime}(\phi) + 3 \frac{\tilde h'(\phi)}{\tilde h(\phi)} \tilde\psi'(\phi) - \omega^2 \tilde\psi(\phi) = - \frac{\tilde J(\phi)}{\tilde h^3(\phi)}.
\label{psi-equation}\end{equation}
Here $\omega^2$ is a combination of the previously introduced exponents:
\begin{equation}
\omega^2 = (\mu + 1)(3\nu - \mu - 1)
\label{omega}\end{equation}
(plotted in Fig.~\ref{omegaeps} as a function of $\alpha$).  Computing $\bfv$ according to prescription~(\ref{vpsi}), we obtain the depth-averaged flow field
\begin{equation}
\bfv = v_r \hat r + v_\phi \hat\phi
\label{vresult}\end{equation}
with components
\begin{equation}
v_r(r,\phi) = - (3\nu - \mu - 1) \frac{J_0}\rho \frac{r^{\mu - \nu}}{\sqrt{A}^{\mu - 1} R^{- \nu + 1}} \tilde h^2(\phi) \tilde\psi(\phi)
\label{vr}\end{equation}
and
\begin{equation}
v_\phi(r,\phi) = \frac{J_0}\rho \frac{r^{\mu - \nu}}{\sqrt{A}^{\mu - 1} R^{- \nu + 1}} \tilde h^2(\phi) \tilde\psi'(\phi).
\label{vphi}\end{equation}
Thus, we need to solve Eq.~(\ref{psi-equation}) with respect to $\tilde\psi(\phi)$ in order to know the flow velocity.

We were not able to find an exact analytical solution to this equation; however, we succeeded in finding approximate solutions on the bisector ($|\phi| \ll \alpha/2$) and near the contact line ($\Delta\phi \ll \alpha/2$), which represent the two opposite limits of the range of $\phi$.  (Again, we define $\Delta\phi = \alpha/2 - |\phi|$.)  Near the contact line the third term on the left-hand side of Eq.~(\ref{psi-equation}) is negligible with respect to the other two terms, and hence the solution is
\begin{equation}
\tilde\psi(\phi) \approx \int_0^\phi \psi^*(\xi) \tilde J(\xi) \, d\xi + \psi^*(\phi) \int_\phi^{\alpha/2} \tilde J(\xi) \, d\xi + \mbox{const}\qquad\qquad(\Delta\phi \ll \alpha/2),
\label{psi-edge}\end{equation}
where
\begin{equation}
\psi^*(\phi) = \int_0^\phi \tilde h^{-3}(\zeta) \, d\zeta.
\end{equation}
The asymptotic of this result is
\begin{equation}
\tilde\psi(\phi) \propto \frac{J^* (\Delta\phi)^{-\lambda-1}}{(1 - \lambda^2) \left|\tilde h'\left(\alpha/2\right)\right|^3}\qquad\qquad(\Delta\phi \to 0).
\label{psiedge}\end{equation}
This asymptotic can also be inferred directly from equation~(\ref{psi-equation}), without finding its solution.  (The constant on the right-hand side of Eq.~(\ref{psi-edge}) turns out to be unimportant compared to the diverging terms.)  In the opposite limit, on the bisector, the second term on the left-hand side of Eq.~(\ref{psi-equation}) is negligible with respect to the other two terms, and therefore the solution is
\begin{equation}
\tilde\psi(\phi) \approx \tilde\psi(0) \cosh(\omega\phi) + \int_0^\phi \frac{\sinh(\omega\xi - \omega\phi) \tilde J(\xi)}{\omega\, \tilde h^3(\xi)} \, d\xi\qquad\qquad(|\phi| \ll \alpha/2).
\label{psi-bisector}\end{equation}
The asymptotic of this result is
\begin{equation}
\tilde\psi(\phi) \propto \tilde\psi(0) + \frac 12 \left(\omega^2 \tilde\psi(0) - \frac{\tilde J(0)}{\tilde h^3(0)}\right) \phi^2\qquad\qquad(\phi \to 0).
\label{psibisector}\end{equation}
Again, this asymptotic can be obtained directly from equation~(\ref{psi-equation}), without solving it.  The reduced pressure on the bisector $\tilde\psi(0)$ is positive, and so is its second derivative $\tilde\psi^{\prime\prime}(0) = \omega^2 \tilde\psi(0) - \tilde J(0) \tilde h^{-3}(0)$.  The latter is due to the facts that $\tilde\psi(\phi)$ diverges {\em positively\/} at the edge of the drop (see Eq.~(\ref{psiedge})) and that ``the Mexican hat'' shape for the pressure as a function of polar angle is physically unlikely in this slow process.  The value of $\tilde\psi(0)$ cannot be determined from the original differential equation; one needs to employ an integral condition resulting from the equality of the total in-flux into a sector of radius $r$ by flow from the outer regions of the drop and the total out-flux from this sector by evaporation:
\begin{equation}
\rho \int_{-\alpha/2}^{\alpha/2} |v_r| h \, r d\phi = \int_0^r \int_{-\alpha/2}^{\alpha/2} J \, r dr d\phi.
\end{equation}
Upon simplification this condition reduces to the following equation defining the constant pre-factor $\tilde\psi(0)$:
\begin{equation}
\int_0^{\alpha/2} \left(\omega^2 \tilde h^3(\phi) \tilde\psi(\phi) - \tilde J(\phi)\right) \, d\phi = 0.
\label{psi0}\end{equation}
Obviously, $\tilde\psi(0)$ is proportional to $\tilde J(0) \tilde h^{-3}(0)$.  Thus, approximate analytical solutions to Eq.~(\ref{psi-equation}) are available in the two opposite limits.

\begin{figure}
\begin{center}
\includegraphics{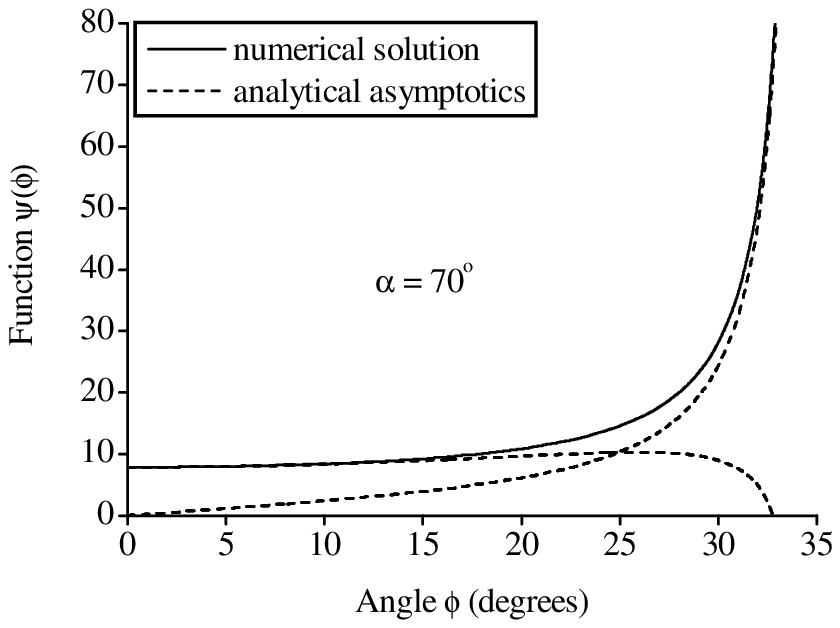}

\includegraphics{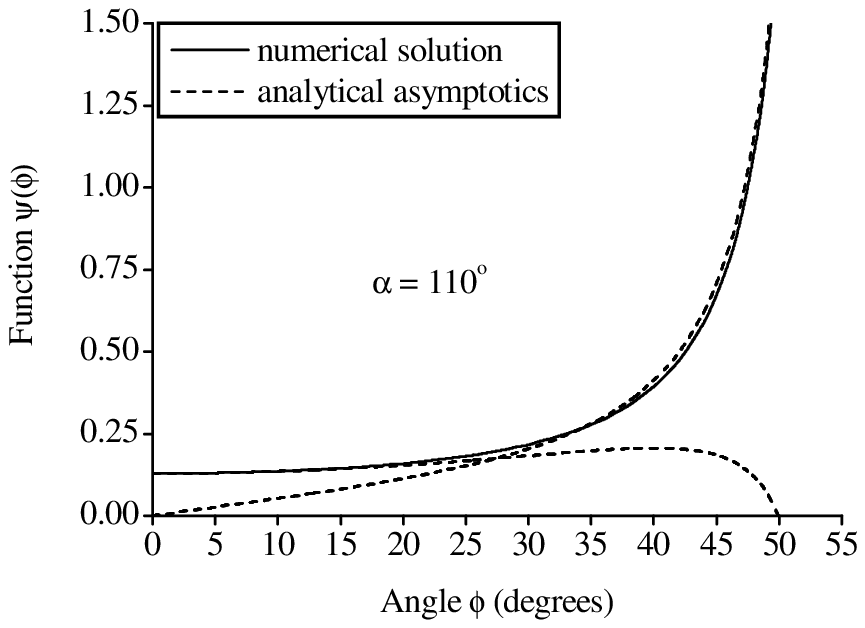}

\caption{Typical behavior of the numerical solution and the analytical asymptotics of function $\tilde\psi(\phi)$ for two values of opening angle.}
\label{psieps}
\end{center}
\end{figure}

In order to compensate for the unavailability of the exact analytical solution to Eq.~(\ref{psi-equation}) we also approached this problem numerically.  The numerical solution to Eq.~(\ref{psi-equation}) satisfying conditions~(\ref{psi0}) for $\tilde\psi(0)$ and $\tilde\psi'(0) = 0$ for $\tilde\psi'(0)$ was found for the two model forms~(\ref{model-2}) and (\ref{model-4}) of $\tilde J(\phi)$ and for approximately 20 different values of the opening angle.  In all cases perfect agreement between the numerical solution and the analytical asymptotics of the previous paragraph was observed.  Two examples of the numerical solution together with the analytical asymptotics are provided in Fig.~\ref{psieps} for opening angles $70^{\circ}$ and $110^{\circ}$.  Both were obtained for the model form~(\ref{model-2}) of function $\tilde J(\phi)$, and the obtuse-angle graph used choice~(\ref{c-choice}) for constant $C$.  Different choice of the model form for function $\tilde J(\phi)$ did not lead to any significant changes of these graphs.

\begin{figure}

\begin{center}
\includegraphics{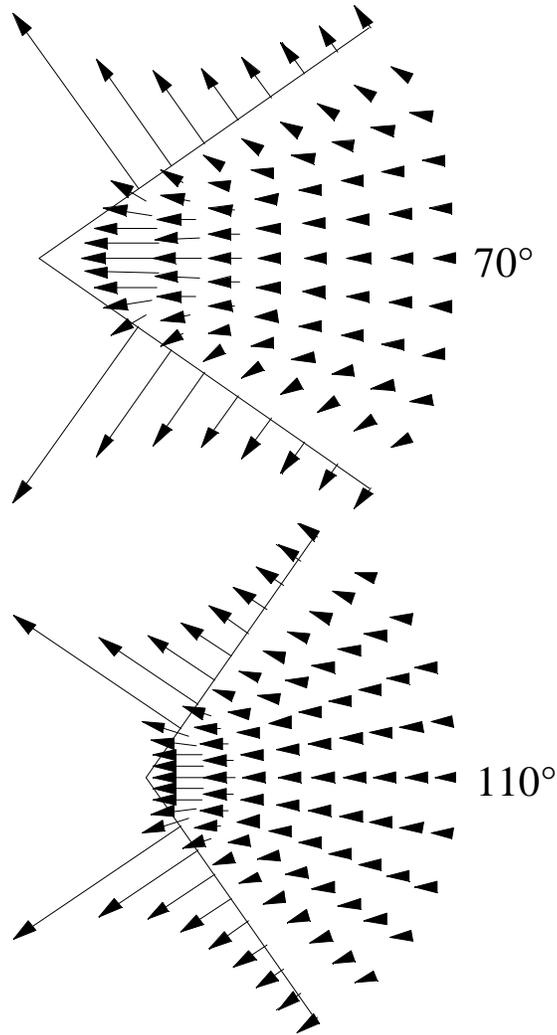}
\caption{Characteristic behavior of flow field for two values of opening angle.  Each arrow represents the absolute value and the direction of velocity $\bfv$ at the point of arrow origin.}
\label{flowfield}
\end{center}
\end{figure}

\begin{figure}
\begin{center}
\includegraphics{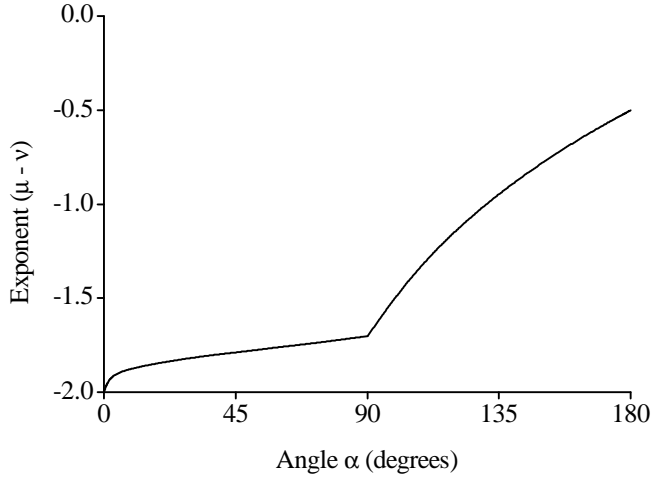}
\caption{Dependence of exponent $(\mu - \nu)$ in the power laws $v_r(r)$ and $v_\phi(r)$ [Eqs.~(\ref{vr}) and (\ref{vphi})] on opening angle $\alpha$.}
\label{munueps}
\end{center}
\end{figure}

Characteristic behavior of the velocity field~(\ref{vresult}) is shown in Fig.~\ref{flowfield} for $\alpha = 70^{\circ}$ and $\alpha = 110^{\circ}$ (again, obtained for the choice~(\ref{c-choice}) and the model form~(\ref{model-2}), but very insensitive to the particular form of $\tilde J(\phi)$).  Note that despite the fact that the exponent $(\mu-\nu)$ of the power law in $r$ is not a smooth function of $\alpha$ (Fig.~\ref{munueps}), the qualitative behavior of the flow field does not visibly change as the opening angle increases past the right angle.

The velocity diverges near the edge of the drop.  This could have been deduced directly from the conservation of mass~(\ref{consmass}), where the divergent evaporation rate must be compensated by the divergent velocity (since the free-surface height is a regular function of coordinates and, moreover, vanishes near the contact line).  Physically, change of volume near the edge becomes increasingly smaller as the contact line is approached and hence the outgoing vapor flux must be matched by an equally strong incoming flow of liquid.

\subparagraph{Streamlines.}  Given the velocity field in the drop, we can now compute the rate of deposit growth at the edge of the sector.  We assume that the suspended particles are carried along by the flow with velocity equal to the fluid velocity.  Integrating the velocity field~(\ref{vr})--(\ref{vphi}):
\begin{equation}
\frac{dr}{r \, d\phi} = \frac{v_r}{v_\phi} = - (3\nu - \mu - 1) \frac{\tilde\psi(\phi)}{\tilde\psi'(\phi)},
\label{velocities-ratio}\end{equation}
we find the streamline equation, {\em i.e.}\ the trajectory of each particle as it moves with the fluid:
\begin{equation}
r(\phi) = r_0 \exp\left[(3\nu - \mu - 1) \int_\phi^{\alpha/2} \frac{\tilde\psi(\xi) \, d\xi}{\tilde\psi'(\xi)}\right],
\label{streamline}\end{equation}
where we assume that $\phi$ is positive here and everywhere below (the generalization to the case of negative $\phi$ is obvious as all functions of $\phi$ are even).  Thus, $r = r_0$ when $\phi = \alpha/2$, so that $r_0$ is the distance from the terminal endpoint of the trajectory to the vertex.  In the limit~(\ref{psiedge}) the integral in the exponent is
\begin{equation}
\int_\phi^{\alpha/2} \frac{\tilde\psi(\xi) \, d\xi}{\tilde\psi'(\xi)} \approx \frac{(\Delta\phi)^2}{2(1 + \lambda)}\qquad\qquad(\Delta\phi \to 0),
\label{integral-early}\end{equation}
and the streamline equation reduces to
\begin{equation}
r \approx r_0\qquad\qquad(\Delta\phi \to 0).
\label{streamline-edge}\end{equation}
The streamlines are perpendicular to the contact line (up to the quadratic terms in $\Delta\phi$).  This is in good agreement with what one would expect near the edge of the drop, since the azimuthal component of the fluid velocity diverges at the side contact line while the radial component goes to zero.  In the limit~(\ref{psibisector}) the integral in the exponent is
\begin{equation}
\int_\phi^{\alpha/2} \frac{\tilde\psi(\xi) \, d\xi}{\tilde\psi'(\xi)} \approx \frac 1{\omega^2 - \kappa^2} \ln\frac{\alpha/2}\phi\qquad\qquad(\phi \to 0),
\label{integral-intermediate}\end{equation}
and hence the result reads
\begin{equation}
r \approx r_0 \left(\frac{\alpha/2}\phi\right)^\epsilon\qquad\qquad(\phi \to 0).
\label{streamline-bisector}\end{equation}
Here we introduced
\begin{equation}
\epsilon = \frac{3\nu - \mu - 1}{\omega^2 - \kappa^2} = \frac 1{\mu + 1 - \kappa^2/(3\nu - \mu - 1)}
\label{epsilon}\end{equation}
and
\begin{equation}
\kappa^2 = \frac{\tilde J(0)}{\tilde h^3(0) \tilde\psi(0)}.
\label{kappa}\end{equation}
A few observations are in order about this limit and its exponents.  First of all, $\kappa^2$ is always positive as all the factors in Eq.~(\ref{kappa}) are.  As we explained in the previous subsection, the second derivative on bisector $\tilde\psi^{\prime\prime}(0) = \omega^2 \tilde\psi(0) - \tilde J(0) \tilde h^{-3}(0)$ has to be positive as well, and therefore $\kappa^2 < \omega^2 = (\mu + 1)(3\nu - \mu - 1)$.  The positiveness of exponent $\epsilon$ follows both from this fact (as $(3\nu - \mu - 1) > 0$ for all $\alpha$) and from the fact that the trajectory $r(\phi)$ necessarily has to diverge as $\phi \to 0$ (as solute comes from the {\em outer\/} regions of the drop).

\begin{figure}
\begin{center}
\includegraphics{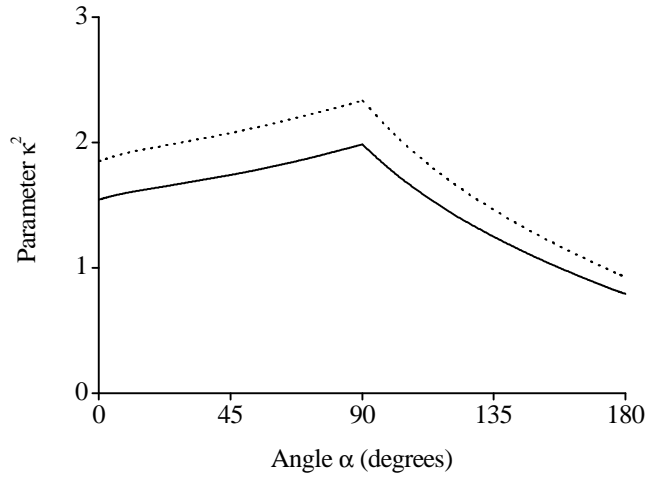}
\caption{Dependence of parameter $\kappa^2$ of Eqs.~(\ref{kappa}) and (\ref{kappa-omega}) [governing exponents $\epsilon$ of Eq.~(\ref{epsilon}), $\gamma$ of Eq.~(\ref{gamma}), and $\delta$ of Eq.~(\ref{delta})] on opening angle $\alpha$.  The two curves correspond to the two model forms for function $\tilde J(\phi)$:  the solid curve is based on the choice~(\ref{model-2}), the dotted curve is based on the choice~(\ref{model-4}).}
\label{kappaeps}
\end{center}
\end{figure}

\begin{figure}
\begin{center}
\includegraphics{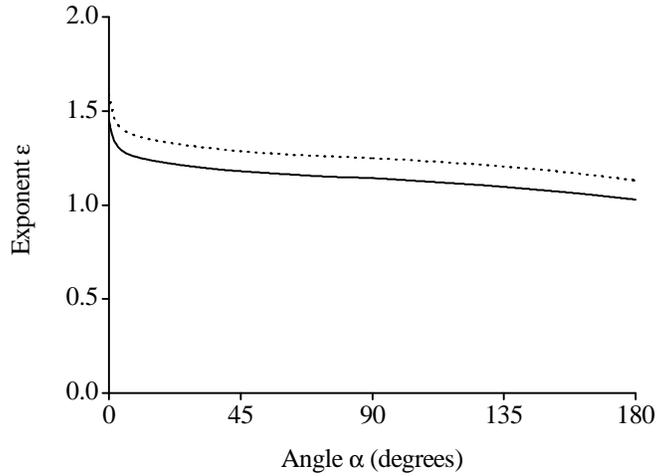}
\caption{Dependence of exponent $\epsilon$ in the power law $r(\phi)$ [Eq.~(\ref{streamline-bisector})] on opening angle $\alpha$.  The two curves correspond to the two model forms for function $\tilde J(\phi)$:  the solid curve is based on the choice~(\ref{model-2}), the dotted curve is based on the choice~(\ref{model-4}).}
\label{epsiloneps}
\end{center}
\end{figure}

We cannot compute $\kappa^2$ and $\epsilon$ explicitly, since we do not know $\tilde\psi(0)$.  However, we can gain some idea of the behavior of these indices by using approximate forms of $\tilde J(\phi)$ and $\tilde\psi(\phi)$.  Fig.~\ref{kappaeps} demonstrates the characteristic behavior of parameter $\kappa^2$ as a function of opening angle, obtained numerically on the basis of the model forms~(\ref{model-2}) and (\ref{model-4}) for function $\tilde J(\phi)$.  Similarly, Fig.~\ref{epsiloneps} shows the behavior of exponent $\epsilon$ for the same two model forms of $\tilde J(\phi)$.  In order to obtain these plots, equation~(\ref{psi-equation}) was solved numerically for each $\alpha$, and then $\tilde\psi(0)$ was fixed by condition~(\ref{psi0}).  As can be observed in these graphs, the two model forms of $\tilde J(\phi)$ lead to the plots of very similar shape, but shifted by approximately 15--20\% for $\kappa^2$ and by no more than 10\% for $\epsilon$ in the whole range of the opening angles.  Thus, we conclude that Figs.~\ref{kappaeps} and \ref{epsiloneps} provide correct estimates for the qualitative behavior and the order of magnitude of parameter $\kappa^2(\alpha)$ and exponent $\epsilon(\alpha)$, respectively.  Interestingly, the exponent $\epsilon$ does not possess a sharp discontinuity of the first derivative at $\alpha = \pi/2$ despite the presence of such discontinuity in parameter $\kappa^2$.

\begin{figure}
\begin{center}
\includegraphics{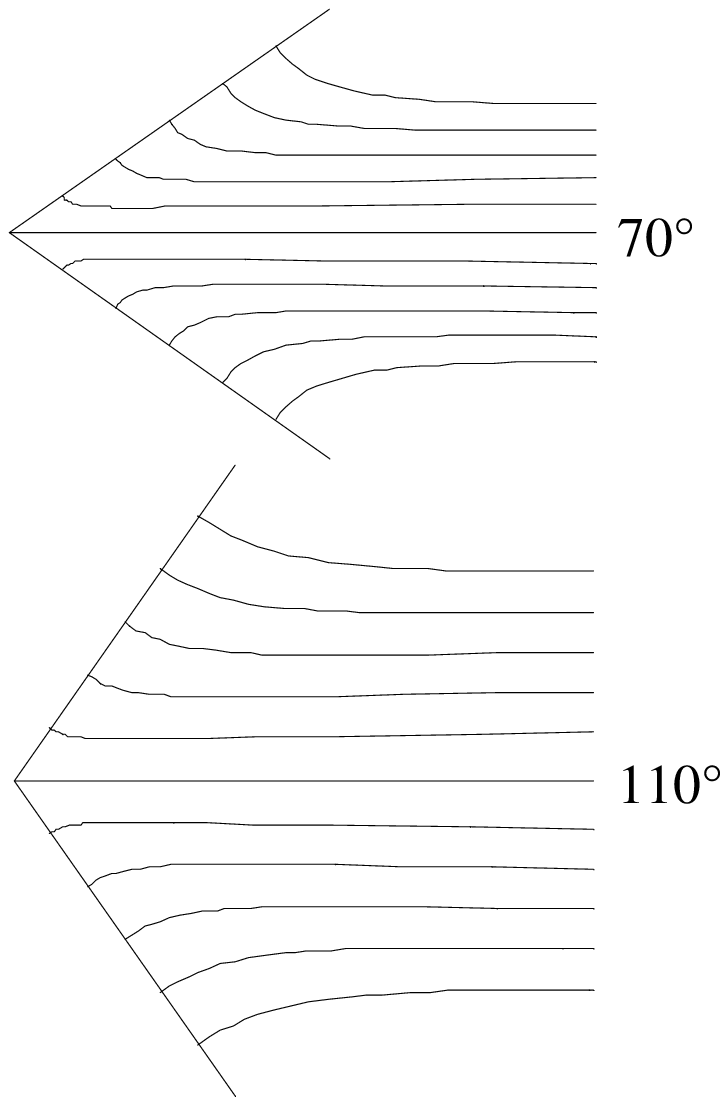}
\caption{Typical shape of the streamlines for two values of opening angle (for the same values as in Figs.~\ref{psieps} and \ref{flowfield}).}
\label{streamlines}
\end{center}
\end{figure}

Typical shape of the streamlines is shown in Fig.~\ref{streamlines} for $\alpha = 70^{\circ}$ and $\alpha = 110^{\circ}$.  It was based on the model form~(\ref{model-2}) for function $\tilde J(\phi)$, and involved the corresponding numerical solutions for function $\tilde\psi(\phi)$ (Fig.~\ref{psieps}) employed in Eq.~(\ref{streamline}).  This shape is practically insensitive to the model form of $\tilde J(\phi)$, and almost an identical copy of this graph was obtained for the model form~(\ref{model-4}).

The distance from a point on a streamline to the bisector scales with $\phi$ as $\phi \, r(\phi) \propto \phi^{1-\epsilon}$ when $\phi \to 0$.  Since $\epsilon > 1$ (Fig.~\ref{epsiloneps}), this distance increases when $\phi$ decreases.  Thus, the streamlines diverge away from the bisector when $\phi \to 0$, and hence they do {\em not\/} originate on the bisector.  An incoming element of fluid initially located close to the bisector moves towards this bisector, reaches a minimum distance and then veers away towards the contact line.  One can also arrive at the same conclusion having started from equation~(\ref{velocities-ratio}).  Indeed, for small $\phi$ (in the limit~(\ref{psibisector})) the ratio of the velocity components~(\ref{velocities-ratio}) is $v_r / v_\phi = - (3\nu - \mu - 1) / (\phi (\omega^2 - \kappa^2))$, or $v_\phi / v_r = - \phi / \epsilon$.  The latter ratio represents the angle between a streamline and a coordinate line $\phi = \phi_0$ at any point $(\phi_0, r(\phi_0))$ on that streamline.  Since $\epsilon > 1$, the absolute value of this angle is less than $|\phi|$, and therefore, despite the opposite sign of this angle, the streamline diverges away from the bisector for small $\phi$.  This tendency can also be observed directly in Fig.~\ref{streamlines}.

Another feature apparent from Fig.~\ref{streamlines} is the self-similarity of all the streamlines.  As is clear from equation~(\ref{streamline}), the only scaling parameter of the family of streamlines is $r_0$, and therefore all the streamlines can be obtained from a single streamline (say, the one with $r_0 = 1$) by multiplying its $r$-coordinate by different values of $r_0$.

Note that equation~(\ref{streamline}) does not contain $J_0/\rho$, and thus it is universally correct regardless of the choice of $J_0$.  Physically, this indicates that solute particles move along the same trajectories independently of how fast evaporation occurs and hence how fast the flow is.  Also, $r(\phi)$ does not depend on the choice of the pre-factor $C$ for the obtuse angles.  Indeed, as we mentioned previously, $\tilde\psi(0)$ is proportional to $\tilde J(0) \tilde h^{-3}(0)$.  Thus, $\tilde\psi(\phi)$ and $\tilde\psi'(\phi)$ are proportional to $1/C^3$ for obtuse angles (this could also be observed directly from equation~(\ref{psi-equation})).  Therefore, the right-hand side of the equation~(\ref{streamline}) is independent of $C$.

These general features of $r(\phi)$ are reflected in the behavior of the index $\kappa^2$ (and hence the exponent $\epsilon$ and all other exponents dependent on $\kappa^2$ that we will introduce later).  Since $\tilde\psi(0)$ is proportional to $\tilde J(0) \tilde h^{-3}(0)$, then $\kappa^2$ is independent of $\tilde J(0) \tilde h^{-3}(0)$ despite the explicit presence of this combination in its definition.  Thus, index $\kappa^2$ is indeed independent of the evaporation intensity and the constant pre-factor of the surface shape, in good agreement with the general observations of the previous paragraph.  On the basis of equation~(\ref{psi0}) defining $\tilde\psi(0)$, index $\kappa^2$ can be written in the form
\begin{equation}
\kappa^2 = \omega^2 \, \frac{\int_0^{\alpha/2} \left(\frac{\tilde h(\phi)}{\tilde h(0)}\right)^3 \frac{\tilde\psi(\phi)}{\tilde\psi(0)} \, d\phi}{\int_0^{\alpha/2} \frac{\tilde J(\phi)}{\tilde J(0)} \, d\phi},
\label{kappa-omega}\end{equation}
demonstrating its independence from the pre-factors of each function of $\phi$.

\begin{figure}
\begin{center}
\includegraphics{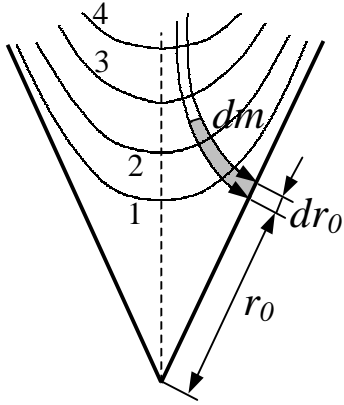}
\caption{Qualitative sketch: mutual location of streamlines (the two lines with arrows) and isochrones (the four numbered lines).  Solute moves along the streamlines towards the contact line (the bold line).  Shaded area is swept by an infinitesimal element of fluid between the two infinitesimally-close streamlines as that element moves towards the contact line.  The isochrones are the geometric locations, starting from which the solute reaches the contact line at the same time.  Solute from isochrone 1 reaches the contact line first; solute from isochrone 4 reaches the contact line last.}
\label{soltrans}
\end{center}
\end{figure}

\subparagraph{Solute transfer.  Three time regimes.}  Now, given the shape of the streamlines, we use our knowledge of the initial distribution of the solute, namely, that the solute has constant concentration $c$ everywhere in the drop at time $t = 0$, and compute the time it takes an element of fluid (moving along a streamline) to reach the contact line at distance $r_0$ from the vertex having started from some point $(r,\phi)$ on that streamline.  This time can be found by integrating either $dt = r \, d\phi/v_\phi$ or $dt = dr/v_r$ with known $v_\phi$ or $v_r$ and the relation between $r$ and $\phi$ on the streamline (Eq.~(\ref{streamline})):
$$t = \int_\phi^{\alpha/2} \frac{r\, d\phi}{v_\phi} = \int_r^{r_0} \frac{dr}{v_r}$$
\begin{equation}
= t_0 \, \int_\phi^{\alpha/2} \frac{\exp\left[(\nu - \mu + 1)(3\nu - \mu - 1) \int_\zeta^{\alpha/2} \frac{\tilde\psi(\xi) \, d\xi}{\tilde\psi'(\xi)}\right]}{\tilde h^2(\zeta) \tilde\psi'(\zeta)} \, d\zeta,
\label{time}\end{equation}
where $t_0$ is a combination of system parameters with dimensionality of time:
\begin{equation}
t_0 = \frac\rho{J_0} \sqrt{A}^{\mu - 1} R^{- \nu + 1} r_0^{\nu - \mu + 1}.
\label{t0}\end{equation}
Within this time {\em all\/} the solute that lays on the way of this element of fluid as it moves toward the contact line becomes part of the deposit (highlighted area in Fig.~\ref{soltrans}).  The mass $dm$ of this deposit (accumulated on the contact line between $r_0$ and $r_0 + dr_0$) can be found by integrating $h(r,\phi)$ over area $dA$ swept by this infinitesimal volume and multiplying the result by the initial concentration $c$ of the solute:
\begin{equation}
dm = c \int_{dA} h(r,\phi) \, r dr d\phi.
\end{equation}
Employing relation~(\ref{streamline}) once again, we obtain:
\begin{equation}
dm = c \, \frac{r_0^{\nu + 1} \, dr_0}{R^{\nu - 1}} \int_\phi^{\alpha/2} \tilde h(\zeta) \exp\left[(\nu + 2)(3\nu - \mu - 1) \int_\zeta^{\alpha/2} \frac{\tilde\psi(\xi) \, d\xi}{\tilde\psi'(\xi)}\right] \, d\zeta.
\label{mass}\end{equation}
Dependence $dm(t)$ can now be found by eliminating $\phi$ from results~(\ref{time}) and (\ref{mass}).  Since we use depth-averaged velocity throughout this paper, we implicitly assume that there is no vertical segregation of the solute.

Exact analytical calculation of the dependence $m(t)$ is not possible for an arbitrary starting point $(r,\phi)$ on a streamline since no analytical expression for $\tilde\psi(\phi)$ is available for arbitrary $\phi$ and since integrals in Eqs.~(\ref{time}) and (\ref{mass}) cannot be computed analytically for arbitrary $\phi$ even if $\tilde\psi(\phi)$ were known.  However, there are two important cases that {\em can\/} be tackled analytically: (a) {\em early times,} when the initial point is close to the contact line ({\em i.e.}\ when $\Delta\phi \ll \alpha/2$ and $r \approx r_0$) and only the solute between that initial point and the contact line is swept into the edge deposit (the starting point is on isochrone 1 of Fig.~\ref{soltrans} or closer to the contact line), and (b) {\em intermediate times,} when the initial point is close to the bisector ({\em i.e.}\ when $|\phi| \ll \alpha/2$ and $r \gg r_0$) and virtually all the solute between the bisector and the contact line is swept into the edge deposit (the starting point is on isochrone 4 of Fig.~\ref{soltrans} or further from the vertex).  Situations between these two limiting cases (highlighted area in Fig.~\ref{soltrans} demonstrates one of them, starting points on isochrones 2 and 3 would correspond to some other) can be extrapolated on the basis of continuity of the results.  Since our region is indefinitely smaller than the drop as a whole, we may treat regions (a) and (b) assuming that a negligible fraction of the drop has evaporated.  At some later stage that we call the {\em late-time\/} regime, an appreciable fraction of the drop has evaporated, and the fluid trajectories have reached back into the bulk of the drop.  In this late regime our asymptotic treatments are clearly not adequate to describe the flow as we did not specify any details of the drop geometry far from the vertex.  Thus we cannot treat this regime by our methods, and only the properties of drying process at early and intermediate stages can be found from information in hand.

Apart from the definitions based on the trajectories, the three regimes can be equivalently defined in terms of time $t$:

\begin{itemize}
\item early times: $t \ll t_0$,
\item intermediate times: $t_0 \ll t \ll t_f$,
\item late times: $t \approx t_f$.
\end{itemize}
Here $t_0$ is the characteristic time defined in Eq.~(\ref{t0}) (this characteristic time depends on $r_0$) and $t_f$ is the total drying time defined in Eq.~(\ref{tf}).  The equivalence of the definitions in terms of the initial position on a trajectory and in terms of time can be seen from equation~(\ref{time}).  At early times, $\Delta\phi \ll \alpha/2$, and the integral in Eq.~(\ref{time}) is much less than 1 (as the integrand is finite near $\phi = \alpha/2$ while the integration range shrinks when $\Delta\phi \to 0$).  Thus, $t \ll t_0$ at early times.  At intermediate times, $|\phi| \ll \alpha/2$, and the integral diverges as $\phi \to 0$.  Hence, $t \gg t_0$ at intermediate times.  On the other hand, the difference between the intermediate and the late regimes lies in their relation to the total drying time as we explained in the previous paragraph.

As is clear from the definition of the intermediate-time regime, the necessary condition for its existence is $t_0 \ll t_f$, which can be reduced to $(r_0/\sqrt{A})^{3 - \mu} (r_0/R)^{\nu - 2} \ll 1$ by combining Eqs.~(\ref{t0}) and (\ref{tf}).  Since we always consider such $r_0$ that $r_0 \ll R$ and $r_0 \ll \sqrt{A}$, this condition is obeyed as long as $3 - \mu > 0$ and $\nu - 2 \ge 0$.  While the former condition is always true ($\mu < 3$ for all $\alpha$), the latter condition is true only for acute opening angles ($\nu = 2$ for $\alpha \le \pi/2$).  Thus, the intermediate-time regime is well-defined for acute angles.  For obtuse opening angles the situation is more complicated.  Index $\nu$ satisfies the opposite inequality ($\nu < 2$ for $\alpha > \pi/2$), and hence for obtuse angles $(r_0/R)^{\nu - 2} \gg 1$ when $r_0 \ll R$.  Combined with inequality $(r_0/\sqrt{A})^{3 - \mu} \ll 1$ this leads to an ambiguous result for how $(r_0/\sqrt{A})^{3 - \mu} (r_0/R)^{\nu - 2}$ compares to 1 and hence how $t_0$ compares to $t_f$.  This result depends on the exact relation between $\sqrt{A}$ and $R$ and on the numerical pre-factor in the definition of $t_f$ that we omitted everywhere (since it depends on the exact shape of the drop including the unspecified regions outside the sector of interest).  Generically, $\sqrt{A} \propto R(t) \tilde\theta(t)$, where $\tilde\theta(t)$ is the contact angle in the bulk of the drop, {\em i.e.}\ far away from the vertex.  Both $R$ and $\tilde\theta$ depend on time; however, the intermediate times are characterized by $t \ll t_f$, and hence, as can be seen from equation~(\ref{r-t}), $R(t) \approx R_i$ and $\tilde\theta(t) \approx \theta_i$ in this regime (here $\theta_i = \tilde\theta(0)$).  Therefore, $\sqrt{A} \propto R_i \theta_i$, and the necessary condition $(r_0/\sqrt{A})^{3 - \mu} (r_0/R)^{\nu - 2} \ll 1$ can be rewritten as $(r_0/\sqrt{A})^{\nu - \mu + 1} \ll \theta_i^{2 - \nu}$.  The exponents on both sides of this inequality are positive for obtuse opening angles, and hence this condition should be expected to be satisfied for not too small initial values of the bulk contact angle $\theta_i$.  The closer to the vertex the trajectory endpoint is, the better this condition is obeyed.  On the other hand, the larger the opening angle is, the smaller $\nu$ is, and hence the smaller $\theta_i^{2 - \nu}$ is (assuming $\theta_i < 1$).  Thus, the condition of applicability of the intermediate-time regime is obeyed worse for larger opening angles.  As we explain below, at exactly $\alpha = \pi$ the intermediate-time regime is indistinguishable from the early-time regime, and hence should not exist.

Results for $m(t,r_0)$ at early and intermediate times are presented in the next two subsections.

\begin{figure}
\begin{center}
\includegraphics{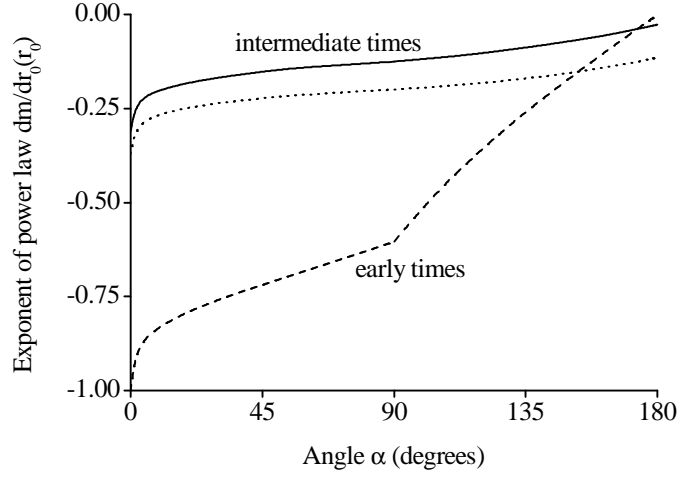}
\caption{Exponent of distance $r_0$ in the power law $dm/dr_0(r_0)$ [Eqs.~(\ref{exponents-early}) and (\ref{exponents-interm})] as a function of the opening angle for the two time regimes.  The early-time curve corresponds to the exponent $\beta$ of Eq.~(\ref{beta}); the intermediate-time curves correspond to the exponent $\gamma$ of Eq.~(\ref{gamma}).  The two curves for the intermediate-time exponent correspond to the two model forms for function $\tilde J(\phi)$:  the solid curve is based on the choice~(\ref{model-2}), the dotted curve is based on the choice~(\ref{model-4}).}
\label{massdist}
\end{center}
\end{figure}

\subparagraph{Deposit growth:  early times.}  This regime corresponds to the growth entirely due to the transfer of particles originally located near the contact line.  The starting point of a particle trajectory is characterized by $r \approx r_0$ and $\Delta\phi \ll \alpha/2$.  An element of fluid $\Delta\phi$ away from the contact line reaches the contact line in time $t$ of Eq.~(\ref{time}).  The mass $dm$ swept to the contact line by an element of fluid $dr_0$-long for this time is defined by Eq.~(\ref{mass}).  In the limit~(\ref{psiedge}) the inner integral in the expressions~(\ref{time}) and (\ref{mass}) is given by equation~(\ref{integral-early}).  Evaluating the outer integrals in these expressions in the limit~(\ref{psiedge}), expressing $\Delta\phi$ in terms of time and then substituting the result into the expression for mass, we finally obtain the mass of the deposit as a function of time:
\begin{equation}
\frac{dm}{dr_0}(t,r_0) \approx c \, \frac{r_0^{\nu + 1}}{R^{\nu - 1}} \frac{\left|\tilde h'\left(\alpha/2\right)\right|}2 \left(\frac{1 + \lambda}{1 - \lambda} \frac{J^*}{\left|\tilde h'\left(\alpha/2\right)\right|} \frac{t}{t_0}\right)^{\frac{2}{1 + \lambda}}.
\label{early-result}\end{equation}
Here $c$ is the constant initial concentration of the solute in the drop.  Note that $t_0$ also depends on $r_0$.  Thus, at early times the deposit grows in time as a power law
\begin{equation}
\frac{dm}{dr_0}(t,r_0) \propto t^{2/(1 + \lambda)} r_0^\beta,
\label{exponents-early}\end{equation}
where the $r_0^\beta$ arises from the $r_0^{\nu + 1}$ pre-factor and from the $r_0$-dependence of $t_0$.  Using Eq.~(\ref{t0}), we find
\begin{equation}
\beta = (\nu + 1) - \frac{2}{1 + \lambda} (\nu - \mu + 1) = - \frac{(1-\lambda)(1+\nu)-2\mu}{1+\lambda}
\label{beta}\end{equation}
and plot it in Fig.~\ref{massdist} as a function of opening angle (the early-time curve).

There are two important conclusions to be drawn from this result.  One is that the power-law exponent of time $2/(1+\lambda) = 4/3$ is exactly the same as in the case of a round drop considered by Deegan {\em et al.}~\cite{deegan1}.  This should be of no surprise since close to the side of the angle (as well as close to the circumference of a round drop) the contact line looks locally like a straight line, and the solute ``does not know'' about the vertex of the angle or the curvature of the circumference.  This exponent is determined entirely by the local properties of an infinitesimal segment of the contact line of length $dr_0$ and is independent of larger geometrical features of the system.

\begin{figure}
\begin{center}
\includegraphics{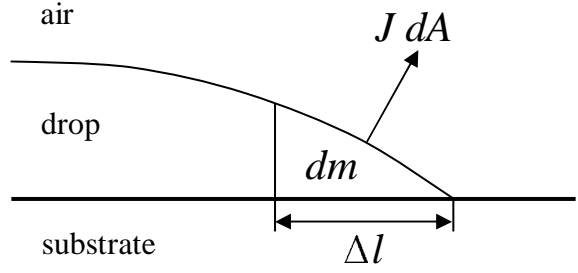}
\caption{An illustration of the derivation of the four-thirds law for $\alpha = \pi$.  The contact line is normal to the plane of the figure.  Length $dr_0$ is along the contact line and hence not shown.  The flow is in the plane of the figure from left to right.}
\label{fourthirds}
\end{center}
\end{figure}

The value $2/(1+\lambda) = 4/3$ of the exponent of time can be obtained from a very simple argument, relying only on the assumptions that (a) the contact line is straight, (b) the streamlines are perpendicular to the contact line, and (c) the distribution of the solute is uniform.  Indeed, the mass of both the water and the solute is proportional to the volume of an element of fluid near the contact line (Fig.~\ref{fourthirds}): $dm \propto (\Delta l)^2 dr_0$.  All this mass should be evaporated from the surface of this volume element in some time $t$.  The evaporation rate (per unit area) scales as $J \propto (\Delta l)^{-\lambda}$ and therefore the rate of mass loss is $J dA \propto (\Delta l)^{-\lambda+1} dr_0$.  The time it takes this volume to evaporate can now be found as the ratio of its mass to the rate of mass loss: $t = dm/(J dA) \propto (\Delta l)^{1+\lambda}$.  Thus, $(\Delta l) \propto t^{1/(1 + \lambda)}$ and hence $dm/dr_0 \propto t^{2/(1 + \lambda)}$ as asserted.

The other observation is the dependence on $r_0$.  Since exponent $\beta$ is always between $-1$ and 0, the singularity in $r_0$ is always integrable at $r_0 = 0$.  Physically, this corresponds to the statement that the vertex of the sector does {\em not\/} dominate the sides and that the deposit accumulation at the vertex is {\em not\/} qualitatively different from the deposit accumulation on the sides.

\subparagraph{Deposit growth:  intermediate times.}  The starting point of a streamline in the intermediate-time regime lies near the bisector and is characterized by coordinates $r \gg r_0$ and $|\phi| \ll \alpha/2$.  By the time an infinitesimal element of fluid from a vicinity of the bisector reaches the contact line virtually {\em all\/} the solute in the area between the bisector and the contact line will be deposited at the contact line.  The analysis is similar to the previous case.  In the limit~(\ref{psibisector}) the inner integral in the expressions~(\ref{time}) and (\ref{mass}) is given by equation~(\ref{integral-intermediate}).  Calculating the time it takes an element of fluid to reach the contact line (Eq.~(\ref{time})) and the mass accumulated at the contact line between $r_0$ and $r_0 + dr_0$ for this time (Eq.~(\ref{mass})), and then eliminating $\phi$ from the two results, we arrive at the dependence of mass on time:
$$\frac{dm}{dr_0}(t,r_0) \approx c \, \frac{r_0^{\nu + 1}}{R^{\nu - 1}} \frac{(\mu + 1)(3\nu - \mu - 1) - \kappa^2}{(\nu - \mu + 1)(3\nu - \mu - 1) + \kappa^2} \,\tilde h(0)\, \frac{\alpha}2$$
\begin{equation}
\times\left((\nu - \mu + 1)(3\nu - \mu - 1)\tilde h^2(0)\tilde\psi(0) \frac{t}{t_0}\right)^{1 + \frac{\kappa^2}{(\nu - \mu + 1)(3\nu - \mu - 1)}}.
\label{interm-result}\end{equation}
Taking into account that $t_0$ also depends on $r_0$, we finally conclude that the deposit mass grows as a power law
\begin{equation}
\frac{dm}{dr_0}(t,r_0) \propto t^\delta r_0^\gamma,
\label{exponents-interm}\end{equation}
where we introduced notations for the exponent of time
\begin{equation}
\delta = 1 + \frac{\kappa^2}{(\nu - \mu + 1)(3\nu - \mu - 1)}
\label{delta}\end{equation}
and for the exponent of $r_0$, originating from both the pre-factor $r_0^{\nu + 1}$ and the $r_0$-dependence of $t_0$ (Eq.~(\ref{t0})):
\begin{equation}
\gamma = (\nu + 1) - \delta (\nu - \mu + 1) = \mu - \frac{\kappa^2}{3\nu - \mu - 1}.
\label{gamma}\end{equation}

An important observation is that the exponent of time stays greater than one in the intermediate-time regime.  Thus, the rate of mass accumulation $dm/dt$ continues to grow with time in this regime, and the deposit mass grows faster and faster.  This result has a simple explanation for both the early- and the intermediate-time regimes.  Since the initial distribution of the solute is uniform, and since the solvent evaporates, the solute concentration at any given volume {\em increases\/} with time.  Thus, even though the fluid and the particles move along the same streamlines in practically constant velocity field (assuming that $R(t) \approx R_i$ at sufficiently early stages), the rate of mass accumulation also {\em increases\/} with time, since portions of solution arriving at the contact line at approximately constant rate have higher and higher solute concentration.  Note that this mechanism and this result are in good agreement with a general conclusion of Deegan's works that the rate of mass accumulation must diverge at the end of the drying process (as $t \to t_f$) and that {\em all\/} the deposit must accumulate at the contact line by $t = t_f$.

Another observation is related to the exponent of $r_0$.  Since $\kappa^2 < \omega^2$ as we showed before, $\gamma > -1$.  Therefore, the mass is integrable at $r_0 = 0$, and the statement of the previous subsection (that the deposit accumulation at the vertex is not qualitatively different from the deposit accumulation on the sides) continues to hold in the intermediate-time regime as well.  Trivially, $\gamma < \mu$.

The exponent of $r_0$ must be identically zero at {\em any\/} time for the opening angle of exactly $\alpha = \pi$.  Indeed, at $\alpha = \pi$ the contact line is just a straight line ({\em i.e.}\ there is no angle at all), and therefore there is a full translational symmetry with respect to which point of this line should be called ``vertex.''  Thus, the choice of $r_0 = 0$ is absolutely arbitrary, and there can be no dependence on $r_0$ whatsoever.

\begin{figure}
\begin{center}
\includegraphics{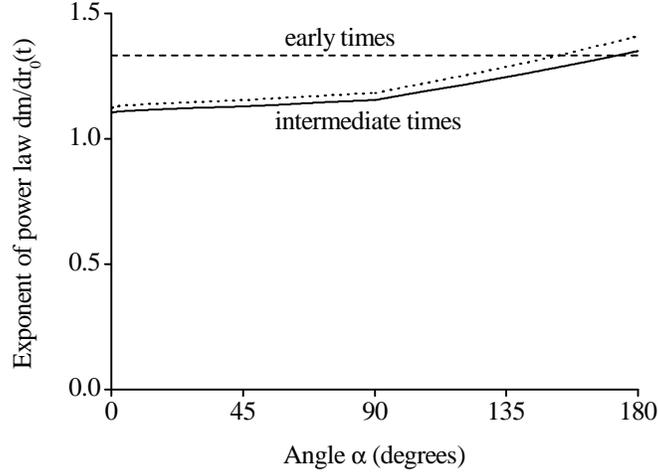}

\caption{Exponent of time $t$ in the power law $dm/dr_0(t)$ [Eqs.~(\ref{exponents-early}) and (\ref{exponents-interm})] as a function of the opening angle for the two time regimes.  The early-time curve corresponds to the exponent $2/(1+\lambda) = 4/3$; the intermediate-time curves correspond to the exponent $\delta$ of Eq.~(\ref{delta}).  The two curves for the intermediate-time exponent correspond to the two model forms for function $\tilde J(\phi)$:  the solid curve is based on the choice~(\ref{model-2}), the dotted curve is based on the choice~(\ref{model-4}).}
\label{masstime}
\end{center}
\end{figure}

Indices $\gamma$ and $\delta$ are plotted in Figs.~\ref{massdist} and \ref{masstime}, respectively, as functions of the opening angle (the intermediate-time curves).  The graphs are based on the result for parameter $\kappa^2$, and the two intermediate-time curves on each graph correspond to the same two model forms~(\ref{model-2}) and (\ref{model-4}) for function $\tilde J(\phi)$ as we used in Fig.~\ref{kappaeps}.  The two model forms of $\tilde J(\phi)$ lead to a very small deviation for $\delta$ (less than 5\%) and to a more substantial difference for $\gamma$.  The significant relative error in exponent $\gamma$ near the value $\alpha = \pi$ is due to the fact that this exponent has to be identically zero at exactly $\alpha = \pi$, while for the model forms of $\tilde J(\phi)$ it is a small, but non-zero number.  Thus, the absolute error is still small, but this small absolute error divided by the small value of the exponent leads to a large relative difference.

\subparagraph{Results for the two time regimes.}  To facilitate the comparison of the results, we plot the exponents for the early- and the intermediate-time regimes in Figs.~\ref{massdist} and \ref{masstime} together.  The absolute values of each exponent are smaller in the intermediate-time regime, indicating that the dependence on distance and time gets {\em weaker\/} with time.  We do not have a simple intuitive explanation for such behavior of these exponents.

The intersection of the exponents near $\alpha = \pi$ on both graphs can be attributed to a couple of reasons.  First, it should be kept in mind that the plotted results for $\gamma$ and $\delta$ are based on the relatively arbitrary choice of model forms~(\ref{model-2}) and (\ref{model-4}) for function $\tilde J(\phi)$.  We suspect that these model equations for the reduced evaporation rate become increasingly inaccurate for large $\alpha$.  For example, $\tilde J(\phi) \propto (\cos\phi)^{-1/2}$ at exactly $\alpha = \pi$, which is not the same as either model form.  Second, as was explained just after Eq.~(\ref{contactangle}), at exactly $\alpha = \pi$ the contact angle $\theta$ is not small even for $r \ll R$, and the correction to the exponent $\lambda$ due to this contact angle (see Eq.~(\ref{lambda})) is comparable to the value $1/2$ assumed in all numerical estimates.  All in all, we believe that this intersection of the early- and intermediate-time exponents is an artifact of our formalism and should not be observed in reality, since the results for the two time regimes should be identical at exactly $\alpha = \pi$.  At exactly $\alpha = \pi$ the contact line is a straight line (no angle) and the trajectories are perpendicular to that straight contact line.  Thus, there should be no differentiation between the early and intermediate times, since this differentiation is based on how far or how close to the bisector the initial point of the trajectory is located, and any perpendicular to a straight line can be called a bisector.  As we showed above, at $\alpha = \pi$ the exponent of $r_0$ must be equal to zero at any time and the exponent of time must be equal to $2/(1+\lambda) = 4/3$ at any time.

\begin{figure}
\begin{center}
\includegraphics{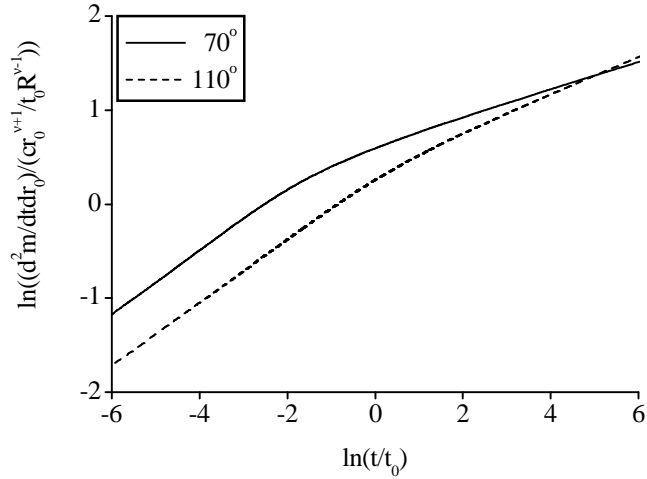}
\caption{Log-log plot of the numerical solution $d^2 m/dtdr_0(t)$ for two values of opening angle (for the same values as in Figs.~\ref{psieps}, \ref{flowfield} and \ref{streamlines}).}
\label{lnmasslntime}
\end{center}
\end{figure}

As we did everywhere above, we also find the numerical solution for $d^2 m/dtdr_0(t)$ in addition to the early- and the intermediate-time analytical asymptotics.  We find the time derivative of $dm/dr_0$ instead of $dm/dr_0$ itself in order to demonstrate the amount of mass arriving at the contact line at time $t$ rather than the total mass accumulated by the time $t$.  We employ the chain rule to obtain $d^2 m/dtdr_0$ on the basis of Eq.~(\ref{time}) for $t(\phi)$ and Eq.~(\ref{mass}) for $dm/dr_0(\phi)$:
\begin{equation}
\frac{d}{dt}\left(\frac{dm}{dr_0}\right) = \frac{\frac{d}{d\phi}\left(\frac{dm}{dr_0}\right)}{\frac{dt}{d\phi}} = \frac{c}{t_0} \, \frac{r_0^{\nu + 1}}{R^{\nu - 1}} \, \tilde h^3(\phi) \tilde\psi'(\phi) \exp\left[(\mu + 1)(3\nu - \mu - 1) \int_\phi^{\alpha/2} \frac{\tilde\psi(\xi) \, d\xi}{\tilde\psi'(\xi)}\right],
\label{dmassdtime}\end{equation}
then use the numerical result for $\tilde\psi(\phi)$ (Fig.~\ref{psieps}) in order to find $t(\phi)$ (Eq.~(\ref{time})) and $d^2 m/dtdr_0(\phi)$ (Eq.~(\ref{dmassdtime})) numerically, and finally create a log-log parametric plot $d^2 m/dtdr_0$ {\em vs.}\ $t$, as shown in Fig.~\ref{lnmasslntime}.  The two curves in Fig.~\ref{lnmasslntime} correspond to the two values of $\alpha$ we used earlier ($70^{\circ}$ and $110^{\circ}$).  Again, the plot is based on the model form~(\ref{model-2}) for function $\tilde J(\phi)$, but very insensitive to the particular form of this function.  This plot clearly demonstrates two different slopes (and hence two different time regimes) of each curve.  The crossover between the two regimes (slopes) occurs around time $t \approx t_0$ ({\em i.e.}\ near $\ln(t/t_0) \approx 0$), and the early-time slopes are equal for both values of the opening angle (and equal to $2/(1+\lambda) - 1 = 1/3$ as to be expected from our early time results).  All these numerical results are in excellent agreement with our analytical predictions, and the numerical values of time exponents compare very well with those of Fig.~\ref{masstime} (which should be corrected by $-1$ due to the differentiation with respect to time in Fig.~\ref{lnmasslntime}).

\begin{figure}

\begin{center}
\includegraphics{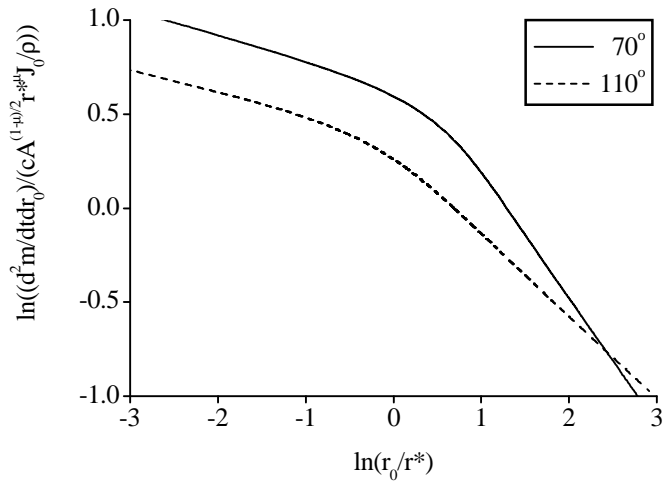}
\caption{Log-log plot of the numerical solution $d^2 m/dtdr_0(r_0)$ for two values of opening angle (for the same values as in Figs.~\ref{psieps}, \ref{flowfield}, \ref{streamlines} and \ref{lnmasslntime}).  Parameter $r^*$ is defined by Eq.~(\ref{r-star}).}
\label{lnmasslndist}
\end{center}
\end{figure}

In a similar fashion we obtain a log-log plot for $d^2 m/dtdr_0$ as a function of $r_0$.  We fix $t$, then express $r_0$ in terms of $\phi$ by combining Eqs.~(\ref{time}) and (\ref{t0}), and finally determine $r_0(\phi)$ and $d^2 m/dtdr_0(\phi)$ (Eq.~(\ref{dmassdtime})) numerically on the basis of the numerical result for $\tilde\psi(\phi)$ of Fig.~\ref{psieps}.  The resulting log-log parametric plot $d^2 m/dtdr_0(r_0)$ is shown in Fig.~\ref{lnmasslndist} for the two values of the opening angle ($70^{\circ}$ and $110^{\circ}$).  The purpose of this graph is to provide a snapshot of the deposit growth at any given moment of time $t$.  For small $r_0$ the accumulation of the solute at the contact line is in the intermediate-time regime, while for large $r_0$ the growth is in the early-time regime.  The threshold between the two regimes is defined by $t = t_0$.  This condition can be reversed by solving Eq.~(\ref{t0}) with respect to $r_0$.  The resulting value
\begin{equation}
r^* = \left( \frac{J_0}\rho \sqrt{A}^{1-\mu} R^{\nu-1} t \right)^{\frac 1{\nu-\mu+1}}
\label{r-star}\end{equation}
defines the threshold in terms of $r_0$ (at any moment of time $t$):  the early regime corresponds to $r_0 \gg r^*$, and the intermediate regime corresponds to $r_0 \ll r^*$.  As can be seen from the numeric plot, the regimes indeed switch at $r_0 \approx r^*$ ({\em i.e.}\ near $\ln(r_0/r^*) \approx 0$).  The intermediate-time slopes are almost equal for both graphs since the intermediate-time exponent $\gamma$ (the upper curves in Fig.~\ref{massdist}) varies very weakly with $\alpha$ ($\gamma \approx -0.135$ for $\alpha = 70^{\circ}$, and $\gamma \approx -0.111$ for $\alpha = 110^{\circ}$).  Again, the numerical results are in excellent agreement with the analytical asymptotics, and the numerical values of exponents compare very well with those of Fig.~\ref{massdist}.

\section{Discussion}

The mechanism presented here does not account for a number of additional effects that can modify the deposition.  First, the vertical distribution of the solute was assumed to be homogeneous throughout the drying process, which is equivalent to assuming that vertical mixing is intensive.  This assumption is quite important, and the results are expected to get modified if the true velocity profile~(\ref{velocity-profile}) is used instead of the vertically-averaged velocity distribution.  Qualitatively, the surface of the drop moves faster than the near-substrate layers, and therefore the particles near the surface reach the contact line sooner than those closer to the substrate.  For example, if all solute particles are confined at the free surface, then the relevant velocity is the surface velocity.  From Eq.~(\ref{velocity-profile}), this surface velocity is $3/2$ times the average velocity $\bfv(r, \phi)$ of Eq.~(\ref{darcy}).  The result is to multiply the characteristic time $t_0$ by $2/3$ in the formulas above.  Additional effects can be caused by gravity or convection, both leading to the non-uniform vertical distribution of solute.

Second, for higher-viscosity liquids viscous stresses become more important, and the typical flow velocities can become comparable with $v^* = \sigma / 3\eta$.  The velocity diverges at the edge, and condition $v \ll v^*$ does get violated at some distance from the edge.  For water this distance is comparable to the typical size of the solute particles (which were 0.1--1~$\mu$m in diameter in the experiments), for higher-viscosity liquids it may become comparable to the size of the drop (a few millimeters).  In the latter case the surface shape cannot be assumed to be equilibrium, as was assumed in this paper.  Also, the characteristics of the flow are significantly altered if the concentration of the solute becomes large (this changes the ``effective'' viscosity of the solvent).

Third, the size of the solute particles can introduce additional effects.   Thus, if the particles are small, diffusion becomes important, and diffusive currents lead to the re-distribution of the solute.  If the particles are too large, sedimentation may affect the amount of the material reaching the edge.  Finally, we ignore the temperature effects on surface tension (Marangoni flows), that have also been observed in the experiments.

Despite all these shortcomings, we believe that our theory captures the essential mechanism of the deposit growth.  The mechanism described is responsible for the transfer of the entire solute to the edge of the drop, and the effects above should not alter our main conclusions.

Our results are similar in many ways to the results for the round drops in works of Deegan {\em et al.}~\cite{deegan1, deegan2}.  For example, the flow is capable of 100\% transfer of the solute to the contact line, there are several distinctive time regimes, the deposit mass grows as a power law of time, and even the exponent of this power law is the same in the early-time regime.  At the same time, there are a number of new features that did not exist in the round-drop case.

One of these features is the existence of the third time regime in addition to the two of the round-drop case.  The intermediate-time regime for the angular sector mathematically corresponds to the late-time regime for the round drop, while the late-time regime for the angular sector does not have any analog in the round-drop case.  The existence of this new regime is due to the fact that an angle, as a mathematical object, is infinite, while a circle always occupies finite area.  A real drop with an angle is also a finite object and hence it must always have a section of the contact line connecting the two sides of the angle.  This causes the existence of the third time regime determined by the solute coming from that far region and influenced by the presence of this new section of the contact line.  The late-time results for the angular sector heavily depend on the shape of this new section.

Probably the most exciting feature of the angular-sector solution is its dependence on the opening angle.  Unlike the round-drop case, there is an extra free parameter of the problem --- the opening angle of the sector.  All the results, including the exponents of the power laws, depend on this opening angle.  Note that these exponents are {\em universal, i.e.}\ they do not depend on any other parameters of the system, except for the opening angle.  They are as universal as the exponent of distance $-2$ in the Coulomb's law.  At the same time, the only parameter they depend on is extremely easy to control --- preparing an evaporating drop one can adjust the opening angle of the contact line at his will without any technical elaborations.  Thus, for example, by suitably choosing the opening angle (and the time regime), one can create a predetermined power-law distribution of the solute along the contact line with virtually any exponent of distance between $-1$ and 0 (Fig.~\ref{massdist}).  In principle, this feature may have significant practical applications for all the processes mentioned in the Introduction.  Further control over the line deposition may be achieved by altering the contact line shape from a straight-sided angle to a curved-sided angle.

Apart from the deposition along a line, a similar virtue of the present flow phenomenon is in setting up a well-defined concentration profile, also depending on $\alpha$.  Initially the amount of solute per unit area is simply proportional to the thickness $h$ of the drop.  As time goes by and the drop dries, the concentration profile changes, and the relative amount of solute per unit area increases as $r$ decreases.  This tendency gets further enhanced with time.  This solute concentration profile may be important if one wants to deposit material {\em within\/} the drop.

While our study does not completely explain the pointed shapes observed in Deegan's experiments (Fig.~\ref{deegan}), it does show that the deposition is strongest near the tip.  Thus, the deposition tends to prevent de-pinning, and it is strongest where the de-pinning force is greatest.  Hence it helps to maintain the angular shapes of Fig.~\ref{deegan}.  Our mechanism does not explain the particular angle observed in the experiments, but, clearly, our kind of analysis is a step towards understanding the chosen angles, the observed separation between the vertices and the deposition profile within the sector.

Our findings have a unique signature that can be readily verified in experiments.  One strong consequence of our theory is that the rate of increase of $dm/dr_0$ has a sharp change of behavior as a function of $r_0$ for any given time (Fig.~\ref{lnmasslndist}).  For small $r_0$ (intermediate times) function $d^2 m/dtdr_0(r_0)$ varies weakly, while for large $r_0$ (early times) it falls off more dramatically with increasing $r_0$.  The crossover point $r^*$ (Eq.~(\ref{r-star})) moves outwards as a power of time $t$ with exponent $1/(\nu-\mu+1)$.  (Note that this exponent involves only the accurately known functions of $\alpha$.)  This crossover point and its outward motion provide a clear-cut signature of our mechanism, and this signature should be the strongest for small opening angles (as Fig.~\ref{massdist} suggests).  In order to avoid possible non-universal effects from late times, one needs to measure the system before the late-time regime.  This measurement can be done by following particles in the flow, or by looking at the build-up of fluorescence at the contact line (both methods were used by Deegan {\em et al.}~\cite{deegan1, deegan2, deegan3, deegan4}).  In order to avoid the uncertainties with depth-averaging, one can employ surface-confined tracer particles.

One open question of our work is related to the unavailability of the exact form of $\tilde J(\phi)$ (as discussed after Eq.~(\ref{theta-def})).  Neither full analytical nor exact numerical form was available explicitly, and we had recourse to analytical asymptotics and approximate numerical expressions.  Finding $\tilde J(\phi)$ may be a formidable task, but can be accomplished at least in principle, as the earlier works on the subject suggest~\cite{kraus, blume1, blume2, desmedt}.  So, one way of determining the exponents more precisely is to try to determine $\tilde J(\phi)$.  Another way is related to creating such evaporating conditions that function $\tilde J(\phi)$ is simpler, for instance, $\tilde J(\phi)$ is just a constant.  The latter case of the uniform evaporation rate is significantly easier to treat analytically, although there are questions on its experimental realization.  Some further efforts may be devoted in this direction.

Further work is also required in order to account for the finite width of the deposition region along the contact line.  It is observed experimentally that the solute is spread out over a quite broad range near the tip and the sides of the angles.  We believe that the finite width of the deposition region is related to the finite size of the solute particles and finite concentration of the solute.  Near the contact line the volume fraction of particles becomes sufficiently high to influence (slow down) the flow that carries those particles.  Further efforts are to be devoted to this problem in the future.

\vspace{3ex}

{\small  The authors acknowledge useful conversations with Todd Dupont.  This work was supported in part by the National Science Foundation under award number DMR-9975533 and in part by its MRSEC Program under award number DMR-0213745.}

\section*{Appendix}

The purpose of this section is to demonstrate that for sufficiently slow flows one can employ the equilibrium surface shape for finding the pressure and the velocity fields instead of having to solve for all the dynamical variables simultaneously.  We will also quantify how slow ``sufficiently slow flows'' are.

We start from the equation of the mechanical equilibrium of the interface~(\ref{mechequil}), where we approximate the doubled mean curvature $2 H$ with $\nabla^2 h$.  This approximation holds true because the free surface of the drop over an angular sector is nearly horizontal near the vertex of the angle, as was shown in our earlier paper~\cite{popov}, and the other terms of the functional $H[h]$ are unimportant.  Substitution of
\begin{equation}
p = - \sigma \nabla^2 h + p_{atm}
\label{p-h}\end{equation}
into the Darcy's law~(\ref{darcy}) yields
\begin{equation}
\bfv = v^* h^2 \nabla (\nabla^2 h),
\label{v-h}\end{equation}
where $v^* = \sigma/3\eta$.  Upon further substitution into the conservation of mass~(\ref{consmass}), we obtain
\begin{equation}
\nabla\cdot\left(v^* h^3 \nabla (\nabla^2 h)\right) + \frac{J}{\rho} + \partial_t h = 0,
\label{consmas1}\end{equation}
which, together with Eq.~(\ref{v-h}), constitutes the full system of equations for finding $h(r,\phi,t)$ and $\bfv(r,\phi,t)$.

Now, for water under normal conditions, $\eta = 1\mbox{ mPa}\cdot\mbox{s}$ and $\sigma = 72\mbox{ mN}/\mbox{m}$.  Hence, the velocity scale $v^*$ is of the order of
\begin{equation}
v^* = \frac\sigma{3\eta} \approx 24\mbox{ m}/\mbox{s}.
\end{equation}
Obviously, this is a huge value compared to the characteristic velocities encountered in usual drying process.  Therefore, one can develop a systematic series expansion in small parameter $\epsilon = \tilde v / v^*$ (where $\tilde v$ is some characteristic value of velocity, say, 10~$\mu$m/s):
\begin{equation}
h = h_0 + \epsilon h_1 + \cdots + \epsilon^n h_n + \cdots,
\end{equation}
\begin{equation}
\bfv = \bfv_0 + \epsilon \bfv_1 + \cdots + \epsilon^n \bfv_n + \cdots,
\end{equation}
and keep only the $h_0$ and $\bfv_0$ terms at the end in order to describe the process up to the main order in $\epsilon = \tilde v / v^*$.  A similar expansion can also be constructed for pressure:
\begin{equation}
p = p_0 + \epsilon p_1 + \cdots + \epsilon^n p_n + \cdots,
\end{equation}
where $p_i$ are related to $h_i$ by equation~(\ref{p-h}):
\begin{equation}
p_0 = - \sigma \nabla^2 h_0 + p_{atm}, \qquad\qquad p_1 = - \sigma \nabla^2 h_1, \qquad\qquad \mbox{etc.}
\end{equation}
Physically, condition $\tilde v \ll v^*$ is equivalent to the statement that the viscous stress is negligible.  Let us understand what $h_0$ and $\bfv_0$ physically correspond to.

Plugging the expansions for $h$ and $\bfv$ into the system~(\ref{v-h})--(\ref{consmas1}), one obtains a set of terms for each power of $\epsilon$, starting from $\epsilon^{-1}$ and up.  Equating terms of the main order in $\epsilon$ yields the following two equations
\begin{equation}
h_0^2 \nabla (\nabla^2 h_0) = 0 \qquad\qquad\mbox{and}\qquad\qquad \nabla\cdot\left(h_0^3 \nabla (\nabla^2 h_0)\right) = 0,
\end{equation}
which both can be satisfied if and only if $\nabla^2 h_0$ is a function of time only.  Writing it as
\begin{equation}
\nabla^2 h_0 = - \frac{p_0 - p_{atm}}\sigma = - \frac 1{R(t)},
\end{equation}
we immediately identify this equation with the statement of spatial constancy of the mean curvature of the interface, which describes the {\em equilibrium\/} surface shape at any given moment of time $t$ ({\em i.e.}\ we obtained equation~(\ref{laplace}) with the desired properties of $p_0$).  Thus, $h_0$ is indeed the equilibrium surface shape, and so is $h$ (up to the corrections of the order of $\tilde v / v^*$).

Repeating the same procedure for the terms of the next order in $\epsilon$, we arrive at another two equations:
\begin{equation}
\bfv_0 = \tilde v h_0^2 \nabla (\nabla^2 h_1),
\end{equation}
\begin{equation}
\tilde v \nabla\cdot\left(h_0^3 \nabla (\nabla^2 h_1)\right) + \frac{J}{\rho} + \partial_t h_0 = 0,
\end{equation}
which can be seen to be equivalent to the set of equations~(\ref{psi})--(\ref{vpsi}) upon identification $\psi = \tilde v \nabla^2 h_1 = - \epsilon p_1 / 3\eta$.  Knowing the equilibrium surface shape $h_0$, one can solve the second equation above with respect to the reduced pressure $\psi$, and then obtain velocity $\bfv_0$ by differentiating the result according to the first equation.  Thus, up to the corrections of the order of $\tilde v / v^*$, one can first find the equilibrium surface shape $h(r,\phi)$ at any given moment of time, and then determine the pressure and the flow fields for this fixed functional form of $h$, as was asserted in section Theory.



\begin{thebibliography}{99}

\bibitem{deegan1} R.D.~Deegan, O.~Bakajin, T.F.~Dupont, G.~Huber, S.R.~Nagel, T.A.~Witten, {\em Phys.\ Rev.\ E\/} {\bf 62}, 756 (2000).

\bibitem{popov} Y.O.~Popov, T.A.~Witten, {\em Eur.\ Phys.\ J.\ E\/} {\bf 6}, 211 (2001).

\bibitem{deegan2} R.D.~Deegan, O.~Bakajin, T.F.~Dupont, G.~Huber, S.R.~Nagel, T.A.~Witten, {\em Nature\/} {\bf 389}, 827 (1997).

\bibitem{deegan3} R.D.~Deegan, {\em Phys.\ Rev.\ E\/} {\bf 61}, 475 (2000).

\bibitem{deegan4} R.D.~Deegan, {\em Ph.D.\ thesis\/} (University of Chicago, Dept.\ of Physics, 1998).

\bibitem{greenspan} H.P.~Greenspan, {\em J.\ Fluid Mech.} {\bf 84}, 125 (1978).

\bibitem{cameron} A.~Cameron, {\em Principles of Lubrication\/} (Longmans, 1966).

\bibitem{brenner} M.P.~Brenner, {\em Ph.D.\ thesis\/} (University of Chicago, Dept.\ of Physics, 1994).

\bibitem{bensimon} D.~Bensimon {\em et al.}, {\em Rev.\ Mod.\ Phys.} {\bf 58}, 977 (1986).

\bibitem{kraus} L.~Kraus, L.M.~Levine, {\em Comm.\ Pure Appl.\ Math.} {\bf 14}, 49 (1961).

\bibitem{blume1} S.~Blume, M.~Kirchner, {\em Optik\/} {\bf 29}, 185 (1969).

\bibitem{blume2} S.~Blume, G.~Kahl, {\em Optik\/} {\bf 70}, 170 (1985).

\bibitem{desmedt} R.~De~Smedt, J.G.~Van~Bladel, {\em IEE Proc.} {\bf 134}, 694 (1987).

\bibitem{jackson} J.D.~Jackson, {\em Classical Electrodynamics,} 2nd edition (Wiley, 1975).

\end{thebibliography}
\end{document}